%% file: access.tex
\documentclass{ieeeaccess}
\usepackage{cite}
\usepackage{amsmath,amssymb,amsfonts}
\usepackage{algorithmic}
\usepackage{graphicx}
\usepackage{textcomp}

\usepackage{placeins}
\usepackage{balance}
\usepackage{hyperref}
\usepackage{url}
\usepackage{subfigure}
\usepackage{algorithm}

\usepackage{booktabs} 
\usepackage{multirow}
\usepackage{framed}
\def\BibTeX{{\rm B\kern-.05em{\sc i\kern-.025em b}\kern-.08em
    T\kern-.1667em\lower.7ex\hbox{E}\kern-.125emX}}
\begin{document}
\history{Date of publication xxxx 00, 0000, date of current version xxxx 00, 0000.}
\doi{10.1109/ACCESS.2023.0322000}

\title{Improving the Representativeness of Simulation Intervals for the Cache Memory System}

\author{\uppercase{Nicolas Bueno}\authorrefmark{1},
\uppercase{Fernando Castro}\authorrefmark{1},
\uppercase{Luis Pinuel}\authorrefmark{1},
\uppercase{Jose I. Gomez-Perez}\authorrefmark{1}
\uppercase{and Francky Catthoor}\authorrefmark{2},\authorrefmark{3}
}

\address[1]{Computer Architecture and Automation Department, Complutense University of Madrid, Madrid, Spain}
\address[2]{Imec, v.z.w. Kapeldreef 75, Leuven, 3001, Belgium}
\address[3]{KU Leuven, ESAT, Oude Markt 13, Leuven, 3000, Belgium}

\tfootnote{This work has been supported by grant PID2021-123041OB-I00 funded by MCIN/AEI/ 10.13039/501100011033 and by “ERDF A way of making Europe”, and by the CM under grant S2018/TCS-4423}

\markboth
{Bueno \headeretal: Improving the Representativeness of Simulation Intervals for the Cache Memory System}
{Bueno \headeretal: Improving the Representativeness of Simulation Intervals for the Cache Memory System}

\corresp{Corresponding author: Fernando Castro (e-mail: fcastror@ucm.es).}

\begin{abstract}
Accurate simulation techniques are indispensable to efficiently propose new memory or architectural organizations. As implementing new hardware concepts in real systems is often not feasible, cycle-accurate simulators employed together with certain benchmarks are commonly used. However, detailed simulators may take too much time to execute these programs until completion. Therefore, several techniques aimed at reducing this time are usually employed. These schemes select fragments of the source code considered as representative of the entire application's behaviour --mainly in terms of performance, but not plenty considering the behaviour of cache memory levels-- and only these intervals are simulated. Our hypothesis is that the different simulation windows currently employed when evaluating microarchitectural proposals, especially those involving the last level cache (LLC), do not reproduce the overall cache behaviour during the entire execution, potentially leading to wrong conclusions on the real performance of the proposals assessed. In this work, we first demonstrate this hypothesis by evaluating different cache replacement policies using various typical simulation approaches. Consequently, we also propose a simulation strategy, based on the applications’ LLC activity, which mimics the overall behaviour of the cache much closer than conventional simulation intervals. Our proposal allows a fairer comparison between cache-related approaches as it reports, on average, a number of changes in the relative order among the policies assessed --with respect to the full simulation-- more than 30\% lower than that of conventional strategies, maintaining the simulation time largely unchanged and without losing accuracy on performance terms, especially for memory-intensive applications.
\end{abstract}


\titlepgskip=-21pt

\maketitle

\input{Intro}

\input{background}
\input{Framework}

\input{motivation.tex}

\input{evaluation}

\input{conclusions}

\bibliographystyle{IEEEtran}
\bibliography{biblio}

\begin{IEEEbiography}[{\includegraphics[width=1in,height=1.25in,clip,keepaspectratio]{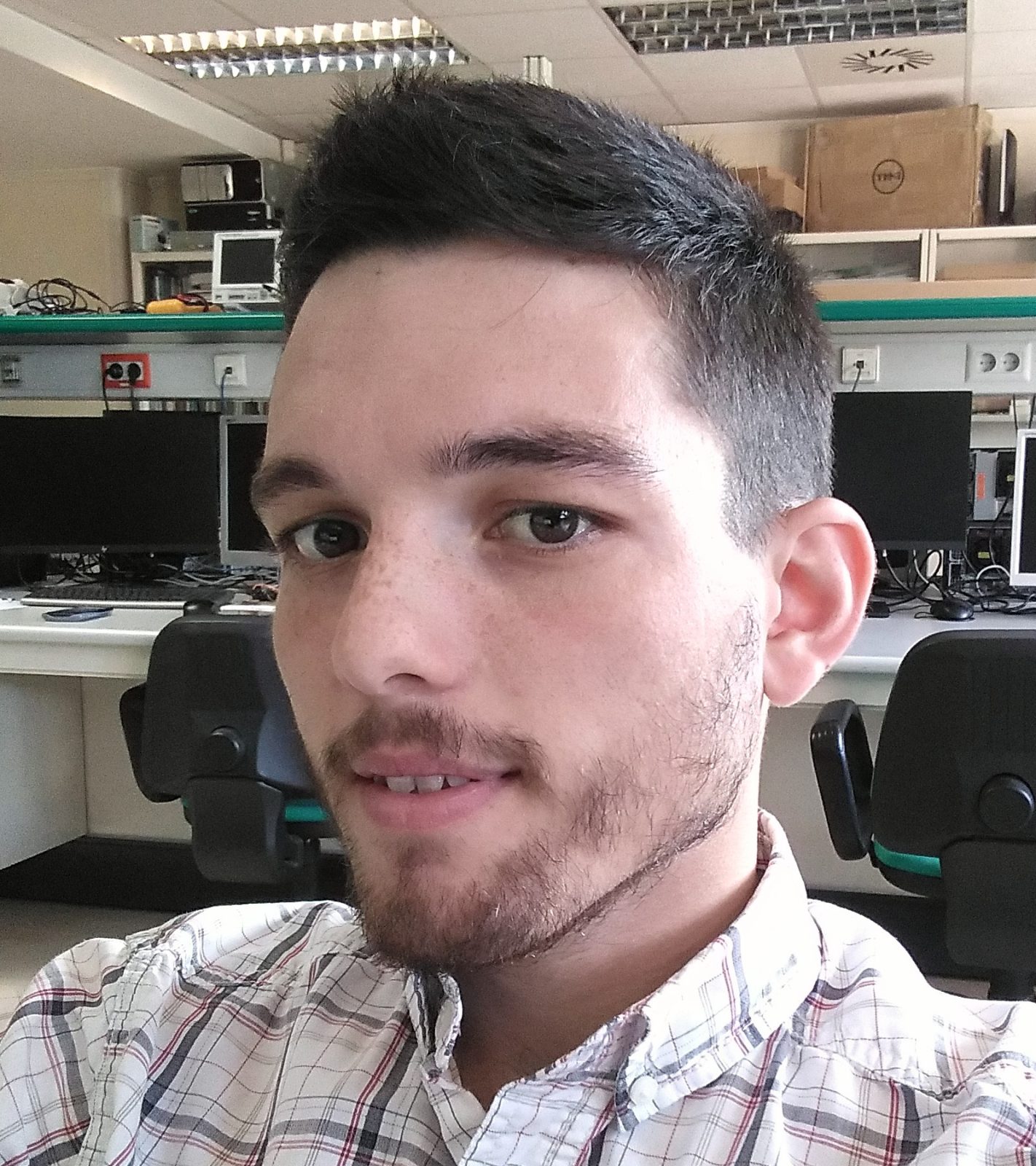}}]{Nicolas Bueno} obtained a Computer Engineering Degree in 2017 and enrolled in a MSc. on Internet of Things by the Complutense University of Madrid. He is currently pursuing the Ph.D. degree in computer engineering at the same university.

His research interests include accurate and fast memory system simulation.
\end{IEEEbiography}

\begin{IEEEbiography}[{\includegraphics[width=1in,height=1.25in,clip,keepaspectratio]{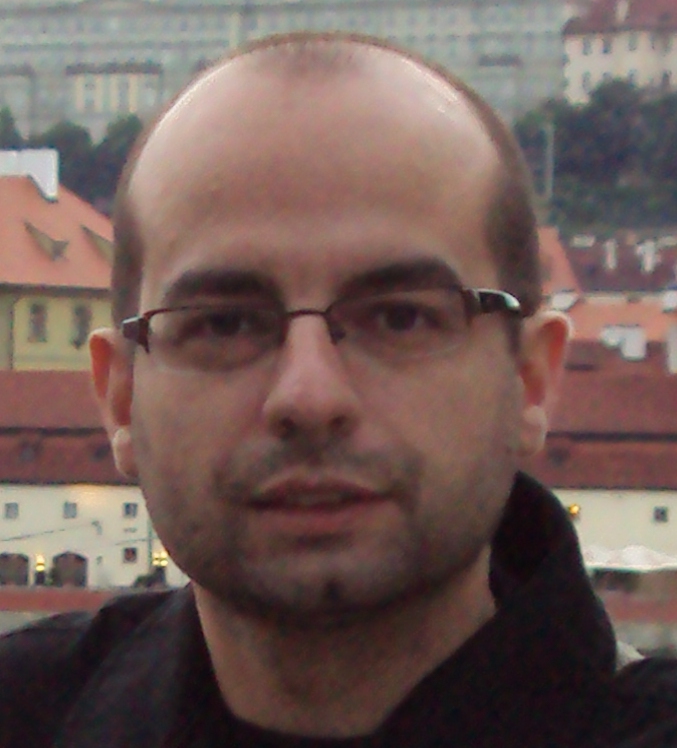}}]{Fernando Castro} obtained the MS degree in Physics from University of Santiago de Compostela (Spain) in 2000, and the MS degree in Electrical Engineering and the Ph.D. degree in Computer Science from the Complutense University of Madrid (Spain) in 2004 and 2008, respectively. 

He currently serves as Associate Professor within the Computer Architecture and Automation department, at Complutense University of Madrid. Since 2003, he continuously participated in competitive projects related to the computer architecture field. He is the author of more than 35 international articles, including publications in international journals with JCR impact factor (as IEEE Transactions on Computers or Journal of Parallel and Distributed Computing) and in the proceedings of very recognized prestige conferences (as IEEE/ACM MICRO, or ACM/IEEE ISLPED). His research interests include energy-aware processor design, efficient memory management (including emerging non-volatile memory technologies) and OS scheduling on heterogeneous multiprocessors.
\end{IEEEbiography}

\begin{IEEEbiography}[{\includegraphics[width=1in,height=1.25in,clip,keepaspectratio]{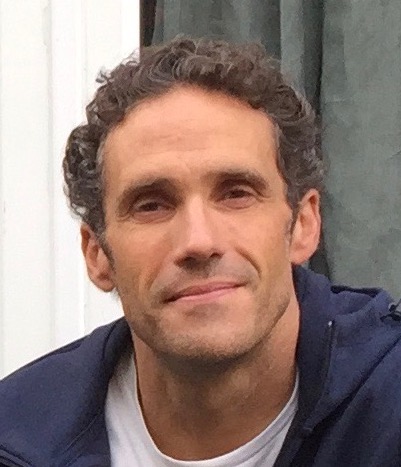}}]{Luis Pinuel} received his M. Sc. and Ph.D. degrees in Computer Science from the Universidad Complutense de Madrid (UCM) in 1996 and 2003, respectively.

He is an Associate Professor of the Department of Computer Architecture and Systems Engineering at the Universidad Complutense de Madrid, Spain. From June 2010 to April 2015 he also served as Academic Secretary of the Physics Faculty at the same University. Previously he was a research assistant at the Acoustic Institute of the Spanish National Research Council (CSIC). His research interests include computer architecture, high-performance computing,low-power microarchitectures, embedded systems, and resource management for emerging computing systems. In these fields, he is co-author of more than 70 publications in prestigious journals and international conferences, several book chapters and he has advised or co-advised 5 PhD dissertations. Ha has been member of the technical program and organization committee of the some relevant conferences (e.g. HPCA). Worried about improving knowledge transfer between research institutions and industry, he has directed more than 15 research contracts with different companies (Texas Instruments, Indra, Satlink, ...).
\end{IEEEbiography}

\begin{IEEEbiography}[{\includegraphics[width=1in,height=1.25in,clip,keepaspectratio]{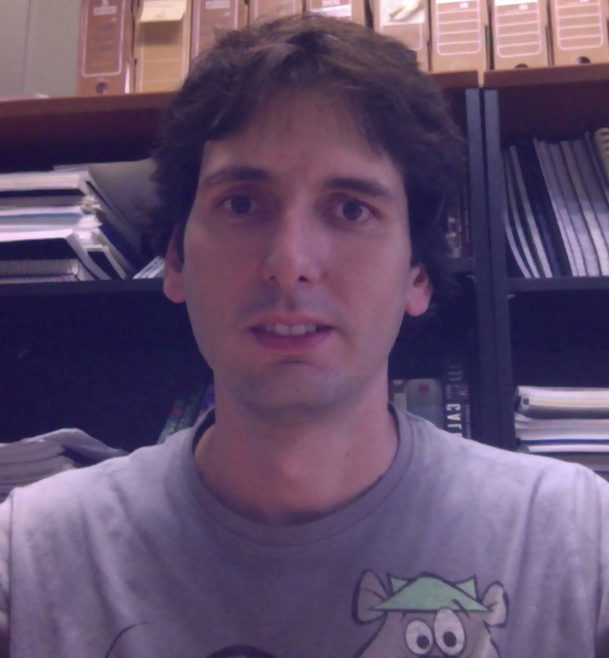}}]{Jose I. Gomez-Perez} received the M.Sc. degree in Computer Science in 2001 and the Ph.D. degree in 2007, all from the Complutense University of Madrid, Spain. 

He is currently an Associate Professor with the Department of Computer Architecture and Automation, UCM, within the ArTeCS group. During his Ph.D., José Ignacio was a visiting researcher at Imec (Leuven),  working on optimizing low-power embedded systems, at both the compiler and system level. After his Ph.D. he moved to GPGPU computing, still at the compiler level. His current research interests include low power embedded systems, focusing on architectural impact of new resistive memory technologies,  in the IoT ecosystem.
\end{IEEEbiography}

\begin{IEEEbiography}[{\includegraphics[width=1in,height=1.25in,clip,keepaspectratio]{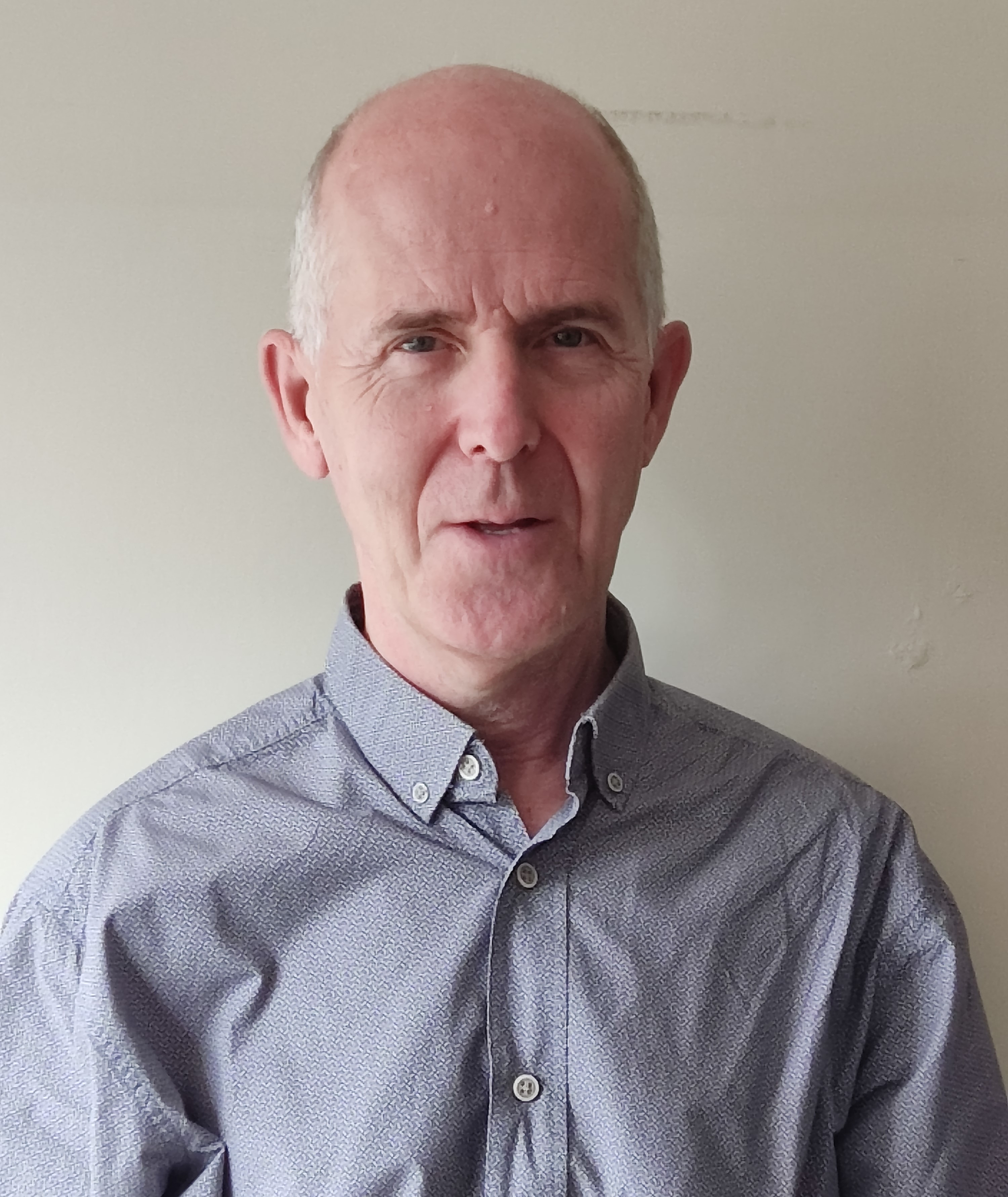}}]{Francky Catthoor} received the Ph.D. degree
in EE from Katholieke University Leuven (KU Leuven), Belgium, in 1987. 

From 1987 to 2000, he headed several research domains in the area of synthesis techniques and architectural methodologies. Since 2000, he has been strongly involving in other activities with IMEC, Leuven, Belgium, including co-exploration of application, computer architecture and deep submicron technology aspects, biomedical systems and IoT sensor nodes, and photo-voltaic modules combined with renewable energy systems. 

Currently, he is an IMEC Senior Fellow. He is also a part-time Full Professor with the EE Department, KU Leuven. He has been an associate editor of several IEEE and ACM journals.
\end{IEEEbiography}

\EOD

\end{document}

%% file: Intro.tex
\section{Introduction}
\label{sec:intro}

\PARstart{C}{urrently}, most researchers in computer architecture employ a real machine or a simulator to evaluate their proposals. However, both approaches exhibit some drawbacks.  

On the one hand, native execution can effectively be employed to evaluate new architectural approaches, but at the cost of a large reduction in exploration space. Fortunately, many commercial systems include performance monitoring support to record execution events and obtain different metrics, that can be used for proposal assessment or benchmark characterization.
 
On the other hand, the entire execution of a benchmark in a cycle-level simulator that models the operation of a complex system (processor, multi-level memory hierarchy, interconnection network, etc.) may require unacceptable long time. Notably, in recent years, processor performance has increased significantly, and also augmented the design complexity of the organizations they currently integrate, mainly multi-core, heterogeneous and specialized-hardware architectures. Moreover, memory hierarchy is one of the principal components because of its significant impact on performance, energy consumption and area occupied, so that an accurate and fast experimental evaluation by means of simulation becomes decisive. Hence, researchers leverage sampling techniques that allow to approximate the behaviour of a full application by using small sections of the program code as simulation intervals~\cite{simflex,simpoint}. However, in microarchitectural research there is great diversity in the selection of these simulation windows. Thus, many authors \cite{hpca_22,hpca_22_2, micro_21, micro_21_2, micro_21_5,isca_21} employ SimPoint \cite{simpoint}, which defines application-specific simulation intervals, 
whereas other authors \cite{nvm_explorer,micro_20,micro_20_2, isca_20} choose to perform an initial fast forwarding or warm up of a determined number of instructions followed by a detailed simulation of a fixed number of subsequent instructions (both processes --forwarding and detailed simulation-- are not application-specific and imply the same number of instructions for all evaluated benchmarks). 
This diversity also exists in the simulator employed (gem5 \cite{gem5} in \cite{hpca_22_2, hpca_22_3, micro_21, micro_21_5,micro_20_2}, Sniper \cite{sniper} in \cite{hpca_22_4,nvm_explorer}, 
or Scarab \cite{scarab} in \cite{micro_21_2,branch_runahead},
among others), the benchmarks used and the input data 
these applications receive (e.g., in the case of SPEC CPU suites, \emph{reference} inputs in
\cite{hpca_22_2,micro_21_5,branch_runahead,hpca_22,Jaleel2010}, \emph{test} inputs in \cite{hpca_22_4} or \emph{train} inputs in \cite{cj_15}). Our motivational hypothesis in this work is that the particular simulation window employed when evaluating microarchitectural proposals related to the last level cache (LLC), such as cache replacement policies, can lead to incorrect conclusions. To demonstrate this, we assess different conventional cache replacement policies using various established simulation intervals, and also simulate the entire benchmarks. The results obtained confirm that the particular simulation window employed significantly affects the relative performance of the policies evaluated. Therefore, we also propose a systematic methodology for selecting simulation intervals aimed at reporting results that reproduce the overall cache behaviour during program execution more accurately than conventional simulation strategies.

To reinforce the demonstration of our hypothesis, in this work we also employ the hardware performance monitoring counters (PMCs) available on a real ARM machine to further study the level of accuracy that SimPoint reports in reproducing the LLC behaviour.  
The motivation behind this combined analysis is that a key aspect in determining the simulation intervals of SimPoint is the correlation between the performance delivered in the complete execution of the benchmark and that obtained when running the selected portions of the program. Nevertheless, the suitability of these simulation intervals for approximating the LLC activity has not been studied in detail previously~\cite{zaragoza_19}. 
Our experiments reveal that, although SimPoint is appropriate for characterizing the entire application behaviour in terms of performance, it fails to properly characterize the LLC behaviour of applications.

\begin{framed}
In this work, we make the following contributions: we demonstrate that 1) following our systematic methodology for using the original SimPoint intervals in a different way --considering applications' LLC activity-- leads to a fairer comparison among cache-related proposals. Also, in the case of memory-intensive programs and compared to conventional simulation strategies, 2) our approach significantly increases the degree of similarity with the full simulation in terms of both cache activity and performance, without impacting on simulation time.
\end{framed}

The rest of the paper is organized as follows: Section \ref{sec:related} presents some background and related work.  
Section \ref{sec:framework} details the experimental framework used.
Section \ref{sec:motivation} motivates and describes our proposed simulation intervals and Section \ref{sec:evaluation} presents the results obtained. 
Finally, Section \ref{sec:conclusions} concludes.

%% file: background.tex
\section{Background and Related Work}
\label{sec:related}

New memory technologies and organizations contribute to significantly augment the complexity in performing an accurate and efficient simulation of the memory hierarchy behaviour. Consequently, many studies have aimed to verify whether current simulation strategies are still valid for new complex memory systems. A widespread strategy consists in characterizing the workloads by selecting a specific subset of benchmarks and then simulating this set using a cycle-accurate simulator~\cite{zaragoza_19,panda_18,limaye_18}.

Next, we briefly describe how SimPoint operates and the cache replacement policies employed in this study. 

\subsection{SimPoint} 
Although it was proposed almost 20 years ago, it is still the most referenced technique for automatic off-line phase detection. 
In selecting the specific fragments of a program code to approximate the behaviour of each full application with a significantly reduced execution time, SimPoint first slices the program into chunks with the same number of instructions. Then, for each chunk, its Basic Block Vector (BBV) is determined, which implies to record the times every single basic block is executed inside the region. After dimensionality reduction performed by random projection, SimPoint employs the K-means algorithm to find the optimal clustering of the program regions, where a similar code is executed and consequently similar behaviour in the system (mainly in terms of performance) is expected. Finally, a single region is selected from each cluster as a representative SimPoint. Although several simulation intervals are determined for each application, only the most representative interval is typically employed in many research works.

\subsection{Cache replacement policies}

In this work, we employ the gem5 simulator, which includes several out-of-the-box replacement policies \cite{cache_policies_gem5}. Each one uses its specific replacement data to determine a replacement victim on evictions \cite{cache_policies_gem5}. Next, we briefly describe the five cache replacement algorithms that we evaluate:

\begin{itemize}
\item{Least Recently Used (LRU)}: The victim is chosen based on a last touch timestamp: the older it is, the more likely its respective entry is to be victimized. 

\item{Tree LRU}: LRU variation that uses a binary tree to keep track of the temporal locality of the entries through 1-bit pointers.

\item{Random}: In this straightforward approach, the block to be replaced is always randomly selected. 

\item{Re-Reference Interval Prediction (RRIP)}: It uses a re-reference prediction value (RRPV) to determine if blocks are going to be re-used in the near future or not. The higher the RRPV value, the more distant the block is from its next access. From the original paper \cite{Jaleel2010}, this implementation of RRIP is also called Static RRIP (SRRIP), as it always inserts blocks with the same RRPV. 

\item{Bimodal Re-Reference Interval Prediction (BRRIP)}: BRRIP \cite{Jaleel2010} modifies the \emph{insertion} of RRIP so that it inserts the majority of cache blocks with a \emph{distant} RRIP and infrequently inserts new blocks with a \emph{long} RRIP.
\end{itemize}

%% file: Framework.tex
\section{Experimental Setup}
\label{sec:framework}

In this section we detail the experimental environments we employed in this work (both a simulator and a real machine) as well as the benchmarks used.

\subsection{Experimental environments}
We motivate and evaluate our proposal by using the gem5 simulator. Moreover, some experiments were conducted on the 2-socket ARM Huawei Taishan 2280 v2 server, equipped with two 64-bit Kunpeng 920 CPUs (model 4826, 48 cores each) running at 2.6 GHz. 
Among all the PMCs available on our platform, we selected those able to measure the events closely related to the cache hierarchy (number of cache misses, cycles executed and instructions retired), employing the perf tool \cite{perf_15} to obtain PMC information. When the gem5 simulator was used, we employed the Syscall Emulation mode and the O3CPU model. Taishan configuration is roughly simulated from available specification data, with the per-core main features of the cache hierarchy shown in Table \ref{tab:cache_param}. Note that no prefetching technique is applied to any cache level.

\subsection{Experimental workloads}
For both the Taishan platform and its simulated gem5 counterpart, we employed the 20 \emph{speed} benchmarks from SPEC CPU2017 suite~\cite{spec_17}, compiled with \emph{gcc v6}. We leverage \emph{train} inputs, using only one input data set per benchmark except in the cases of \emph{perlbench} (5), \emph{gcc} (3), \emph{bwaves} (2), \emph{xz} (2) and \emph{nab} (2), where we experiment with different inputs (the particular number is expressed in parentheses right after the name of the benchmark). 

\begin{table}[b]
\caption{Cache parameters employed in the gem5 simulator\label{tab:cache_param}}
\begin{center}{\tt
    \begin{tabular}{|l|c|c|c|c|c|}
    \hline
{}&\textbf{Type}&\textbf{Assoc.}&\textbf{Size}&\textbf{Latency}&\textbf{MSHR}\\
    \hline
    DL1/IL1 & Private & 4 & 64KB & 4 cycles & 4\\
    \hline
    L2 & Private & 8 & 512KB & 8 cycles & 20\\
    \hline
    L3 (LLC) & Shared & 8 & 1MB & 37 cycles & 24\\
    \hline
    \end{tabular}}
\end{center}
\end{table}

%% file: motivation.tex
\section{Motivation and Proposal}
\label{sec:motivation}

To motivate our work, we first evaluate with gem5
--using the settings detailed in Table \ref{tab:cache_param}-- 
the five aforementioned cache replacement policies by running the \emph{speed} benchmarks from the SPEC CPU2017 suite under four different simulation approaches:

\begin{itemize}
\item {Fast-forwarding of the first 1000M instructions followed by a detailed simulation of the subsequent 1000M or 2000M instructions (we refer to these simulation strategies as \emph{ff1000} and \emph{ff2000}, respectively)}.
\item {Simulation of 100M-instruction windows according to SimPoint (we denote this strategy as \emph{spt}). Note that, 
for the sake of fairness, we employ \textbf{all} the SimPoint simulation intervals, not only that of the highest weight. Table \ref{tab:number_simpoints} shows the specific number of simulation intervals employed for each application used. It is also worth noting that as we are using windows of enough interval size, no warmup is required, as other works pointed out \cite{warmup_1,warmup_2,warmup_3}}.   
\item {Full simulation. We drop intermediate results every 100 ms so that we can partly reconstruct the temporal behaviour of the application (we refer to this approach as \emph{full})}.
\end{itemize}

\subsection{Motivational analysis}
\label{subsec:mot}

In this section we aim to validate our hypothesis. Recall that it states that \emph{the chosen simulation intervals may lead to incorrect conclusions when exploring cache-related microarchitectural proposals}.

\begin{figure*}[htbp]
\centering
\subfigure[pop2]{\includegraphics[width=58mm]{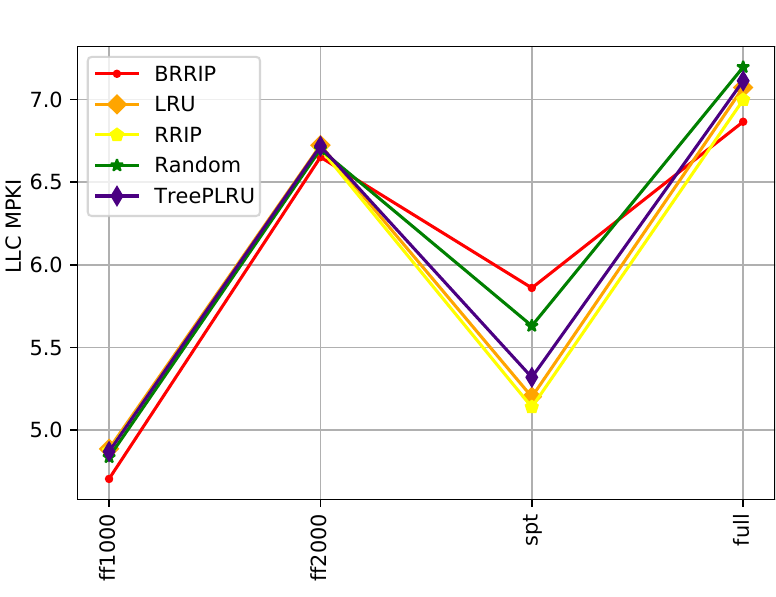}}
\subfigure[gcc with train input]{\includegraphics[width=58mm]{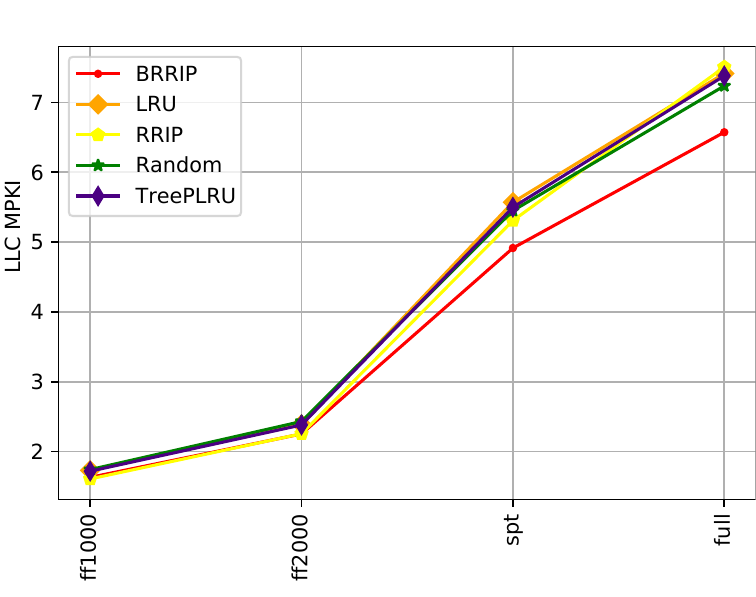}}
\subfigure[mcf]{\includegraphics[width=58mm]{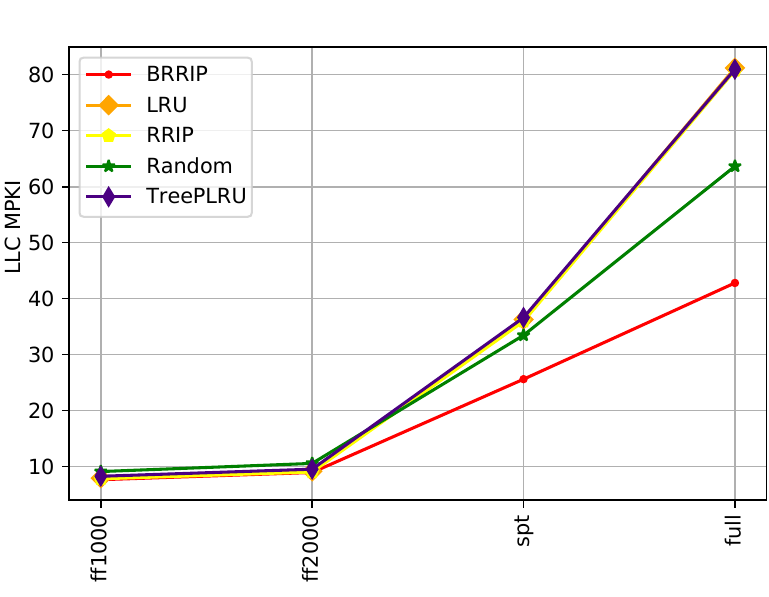}}
\subfigure[x264]{\includegraphics[width=58mm]{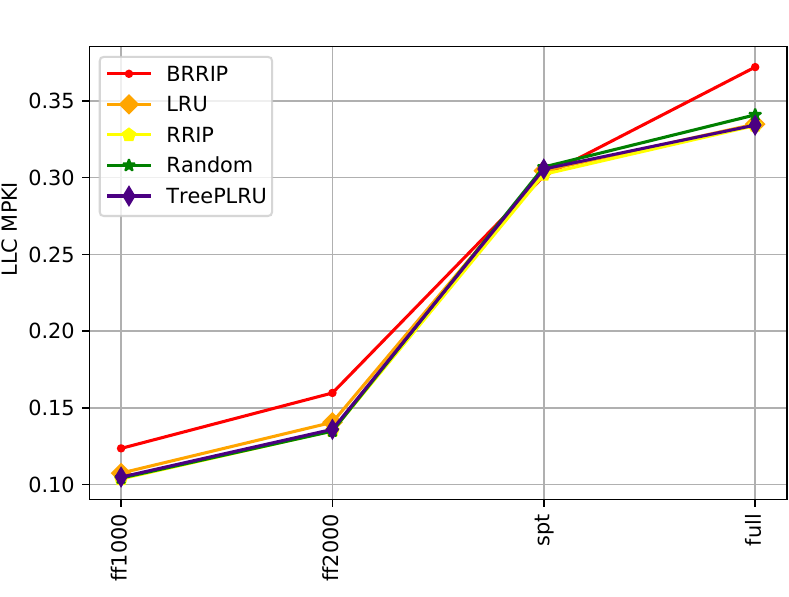}}
\subfigure[imagick]{\includegraphics[width=58mm]{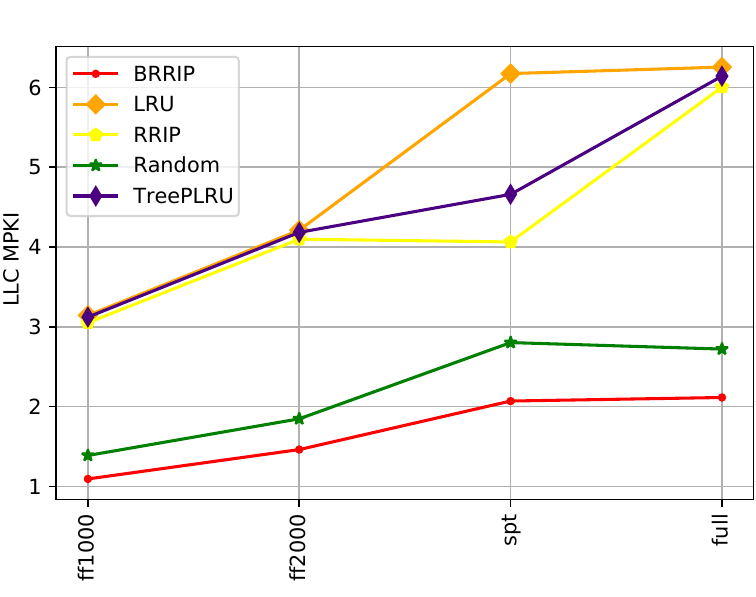}}
\subfigure[roms]{\includegraphics[width=58mm]{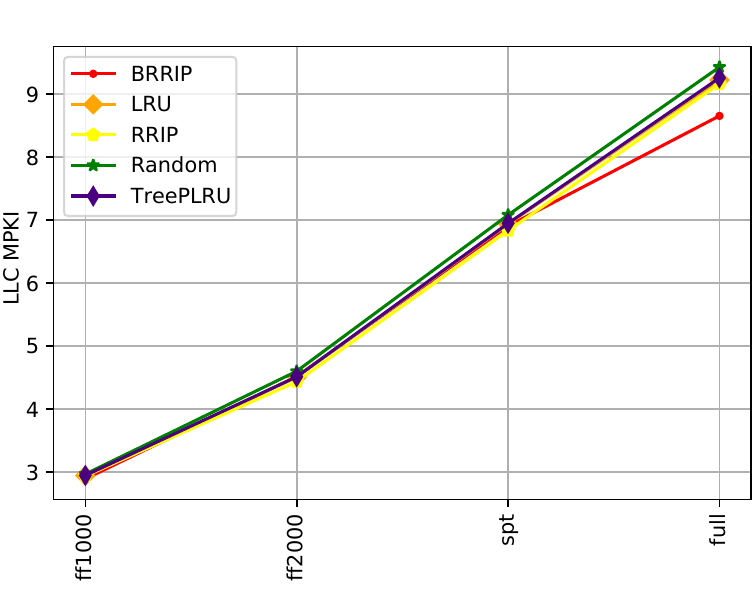}}
\caption{LLC MPKI obtained with the four simulation strategies, using gem5, for different benchmarks and cache replacement policies.} \label{fig:mpkis}
\end{figure*}

\subsubsection{General behaviour}

We explore the LLC misses per 1K instructions (MPKI) and the cycles per instruction (CPI) values for the evaluated applications. 

Regarding the LLC misses it is worth noting that we measure the total numbers of misses, including both data misses and instruction misses without distinction. This is based on the fact that instruction misses in the LLC are extremely rare. Recall that our simulated configuration roughly models that of the Taishan platform, which features a separated first level cache for instructions (IL1) and data (DL1), whereas the second level cache (L2) and the LLC (L3) are both shared by instructions and data, as typically occurs in commodity systems. According to our experiments, instruction misses in LLC represents less than 0.5\% of the total LLC misses for the vast majority of the benchmarks assessed. The average value, using the arithmetic mean and considering all the 20 applications, is around 5.5\%. If we omit the contribution of the very few outlier benchmarks exhibiting a high percentage of LLC instruction misses, this number is just 0.8\%. If we employ the geometric mean, the average value is 0.06\% considering all the 20 applications and 0.03\% removing the contribution of the outliers.  

As a representative sample of the results obtained, Fig. \ref{fig:mpkis} shows the MPKI values reported in the execution of six benchmarks under the four different simulation strategies described (the y-axis represents the absolute number of MPKI, so the scales in each figure may be different). The following conclusions can be drawn:

\begin{table}
    \caption{Number of SimPoint intervals per evaluated benchmark-input pairs and LLC MPKI values obtained with LRU policy}
    \centering
    \begin{tabular}{|c|c|c|c|}
    \hline
     \textbf{Benchmark}   & \textbf{Input} & \textbf{\# SimPoints} & \textbf{MPKI (LRU)}\\
     \hline
     \multirow{5}{4em}{perlbench }  & diffmail & 12 & 0.31\\
     \cline{2-4}
         & perfect & 6 & 0.01\\
         \cline{2-4}
         & scrabbl & 9 & 0.16\\
         \cline{2-4}
         & splitmail & 13 & 0.80\\
         \cline{2-4}
         & suns & 15 & 1.65 \\
    \hline
    \multirow{3}{2em}{gcc}   & 200 & 19 & 5.01 \\
    \cline{2-4}
         & scilab & 21 & 3.55\\
         \cline{2-4}
         & train & 7 & 7.25\\
    \hline
    \multirow{2}{3.5em}{bwaves}   & bwaves1 & 16 & 9.03 \\
    \cline{2-4}
         & bwaves2 & 14 & 8.52\\
    \hline
    mcf    & train & 23 & 81.14\\
    \hline
    cactuBSSN    & train & 20 & 6.62\\
   \hline
   lbm    & train & 27 & 41.09\\
   \hline
   omnetpp    & train & 25 & 11.43\\
   \hline
   wrf    & train & 25 & 12.52\\
   \hline
   xalancbmk    & train & 27 & 0.08\\
   \hline
   x264    & train & 13 & 0.34\\
   \hline
   cam4    & train & 28 & 4.91\\
   \hline
   pop2    & train & 20 & 7.11\\
   \hline
   deepsjeng    & train & 12 & 0.26\\
   \hline
   imagick    & train & 22 & 6.23\\
   \hline
   leela    & train & 16 & 0.38\\
    \hline
    \multirow{2}{1.75em}{nab }   & aminos & 23 & 3x$10^{-4}$ \\
    \cline{2-4}
         & gcn4dna & 17 & 0.47\\
    \hline
    exchange2    & train & 16 & $10^{-5}$\\
    \hline
    fotonik3d    & train & 21 & 25.11 \\
    \hline
    roms    & train & 22 & 9.21\\
    \hline
    \multirow{2}{1.5em}{xz }   & input\_combined 40 & 16 & 1.91 \\
    \cline{2-4}
         & IMG\_2560 40 & 16 & 5.82\\
    \hline
    \end{tabular}
    \label{tab:number_simpoints}
\end{table}

\begin{figure*}[t]
\centering
\subfigure[pop2]{\includegraphics[width=58mm]{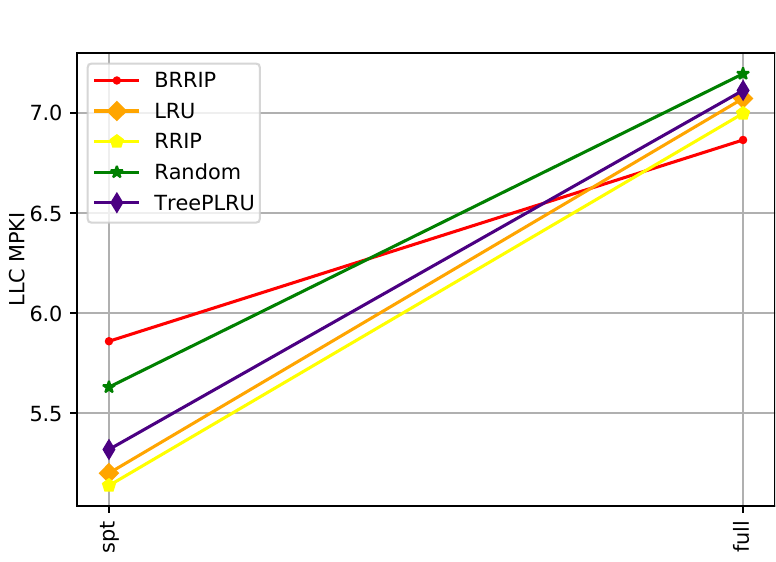}}
\subfigure[gcc with train input]{\includegraphics[width=58mm]{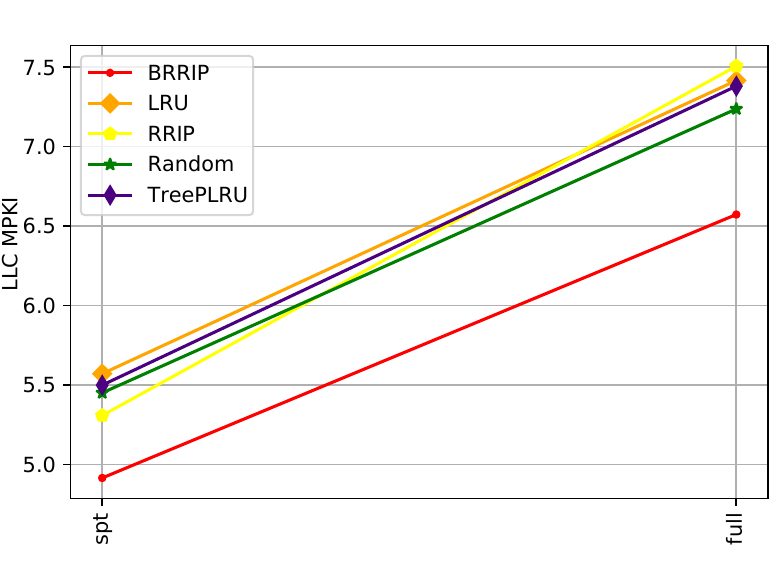}}
\subfigure[mcf]{\includegraphics[width=58mm]{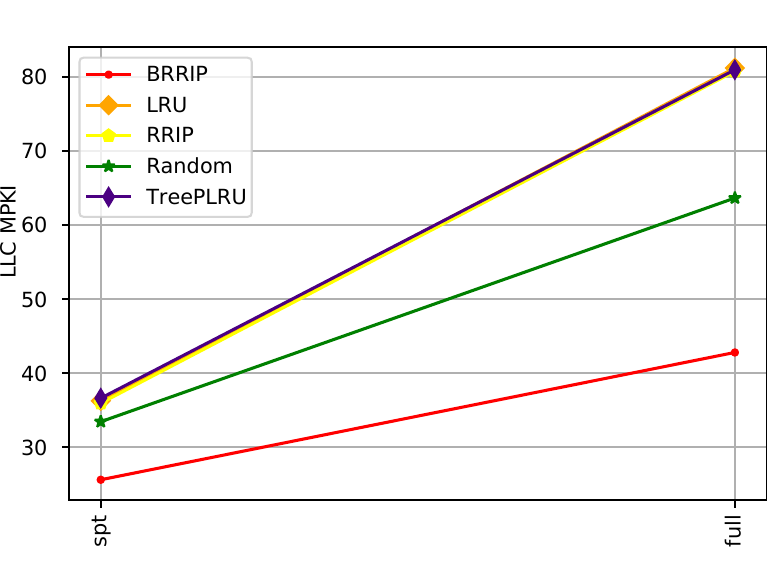}}
\subfigure[x264]{\includegraphics[width=58mm]{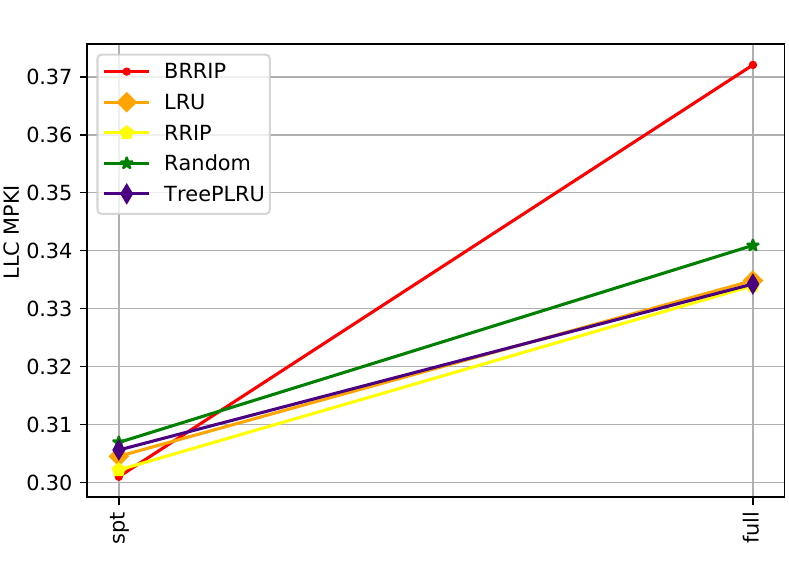}}
\subfigure[imagick]{\includegraphics[width=58mm]{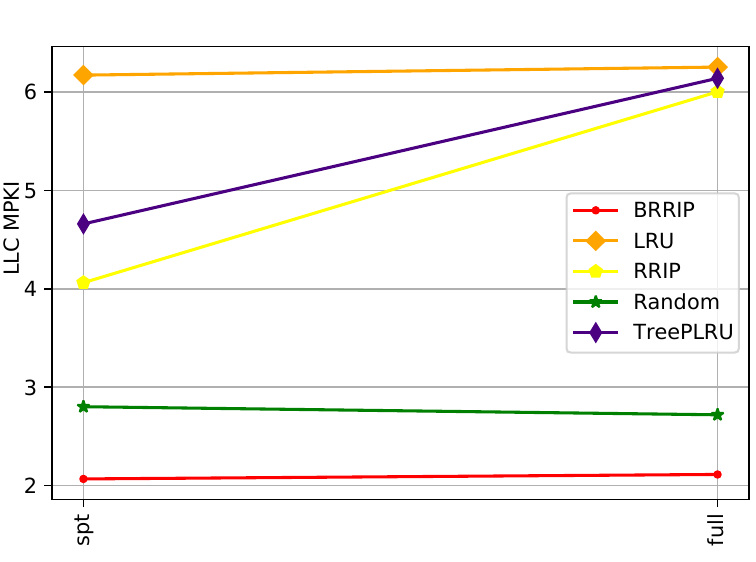}}
\subfigure[roms]{\includegraphics[width=58mm]{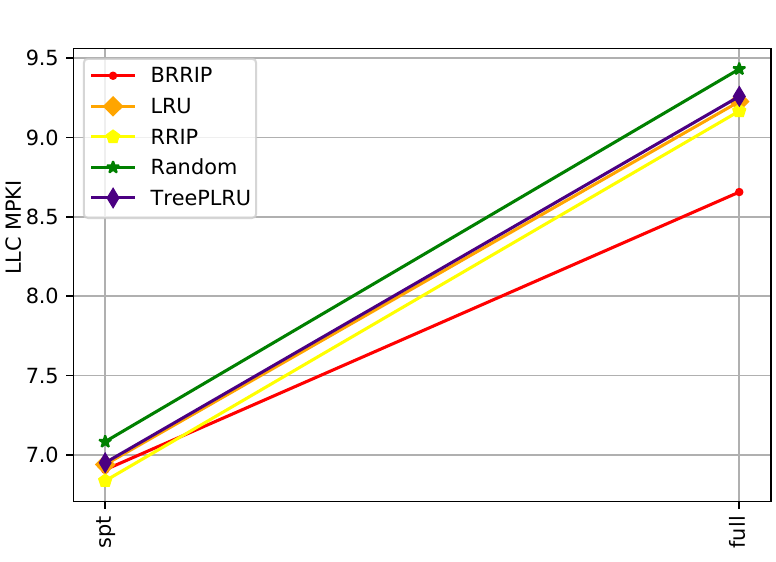}}
\caption{Zoom into LLC MPKI obtained with SimPoint and full simulation, using gem5, for different benchmarks and cache replacement policies.} \label{fig:mpkis_zoom}
\end{figure*}

First, for all the applications shown, we can observe how the MPKI values vary significantly depending on the specific intervals simulated. MPKI figures obtained with the full simulation are generally substantially higher than those reported by the first two simulation windows. When using \emph{fast-forwarding}, doubling the simulation window slightly improves the results, but they generally remain significantly far from the reference (except for \emph{pop2}). Therefore, the results obtained when using these two first simulation strategies seems to be not representative of the application's overall behaviour, which suggests that no reliable conclusions on the cache-related proposals evaluated under these strategies can be extracted.
SimPoint results are closer to the full simulation results (except for \emph{pop2} again), but the differences, in relative terms, are also notable.  
In the applications shown, these variations range up to 20-55\% for some cache replacement policies. 
In the case of \emph{mcf} --the benchmark with the highest LLC MPKI among the 20 evaluated applications, as shown in Table \ref{tab:number_simpoints}-- LRU, Tree LRU and RRIP policies all report an MPKI value with SimPoint approximately 55\% lower than when using full simulation. These significant differences also occur in most of the evaluated applications.  
Actually, considering the maximum difference (between MPKI values using SimPoint and the full simulation) reported by any of the replacement policies assessed for each benchmark, the average value considering all benchmarks is around 23\%. Note that each benchmark contributes only one value to the final mean (we use the average value for benchmarks with more than one input). In addition, we excluded the contribution of \emph{xalancbmk}, \emph{exchange2}, \emph{perlbench} (\emph{perfect} input) and \emph{nab} (\emph{aminos} input), because they all report very low MPKI values (below 0.1 --we choose this threshold as it constitutes the 1\% of the average MPKI obtained with the LRU policy considering the 20 benchmarks evaluated--), so that relatively small variations in absolute numbers lead to extraordinarily high variations in percentage, distorting the final result. Note that all the numbers reported in this experiment were obtained with the same simulator toolchain, so comparisons between the different approaches are fair.

\begin{framed}
Even when considering simulator inaccuracies, we believe that these large relative differences are representative, and both \emph{fast-forwarding} and SimPoint consistently underestimate the MPKI in the LLC, worsening the representativeness of the simulation intervals for the cache system with respect to the entire application’s behaviour.
\end{framed}

Second, and more importantly, traditional simulation techniques may lead to incorrect conclusions regarding the relative efficiency of the different replacement policies. 
For clarity, Fig. \ref{fig:mpkis_zoom} zooms into the MPKI results reported by the corresponding SimPoint intervals and the full simulation of the same six benchmarks shown in Fig. \ref{fig:mpkis}. 
According to these data, the relative order between different policies is changed in four (\emph{pop2}, \emph{gcc}, \emph{x264} and \emph{roms}) of these six cases. Thus, for example, if we were to compare RRIP with LRU policies in \emph{gcc}, the selected SimPoint intervals would benefit the former, whereas LRU performs better when considering the whole benchmark. 
This benchmark and \emph{x264} may not be statistically significant, because the absolute values of the differences are quite small. \emph{pop2} shows a more clear behaviour in that concern, with BRRIP clearly penalized when using SimPoint, behaviour also observed in \emph{roms}. 
In the case of \emph{mcf} and \emph{imagick} (and other applications not shown), although no changes in the relative order between the different replacement policies is observed, we can also note significant relative variations. For example, in \emph{mcf}, whereas using SimPoint the BRRIP approach reports an MPKI value around 27\% lower than those of LRU, TreeLRU and RRIP policies, this percentage rises to 47\% in the full simulation.
Regarding the rest of applications evaluated, we also obtain significant changes in the relative order between the replacement policies employed depending on the simulation strategy used (in particular when comparing SimPoint and the full simulation) in all the benchmarks evaluated except \emph{bwaves}, \emph{cactuBSSN}, \emph{wrf}, \emph{deepsjeng}, \emph{nab}, \emph{exchange2}, \emph{fotonik}, \emph{xz} and \emph{lbm}, so we can conclude that these changes occur in 
roughly a half of the evaluated benchmarks. 

\begin{figure*}[t]
\centering
\subfigure[perlbench with splitmail input]{\includegraphics[width=58mm]{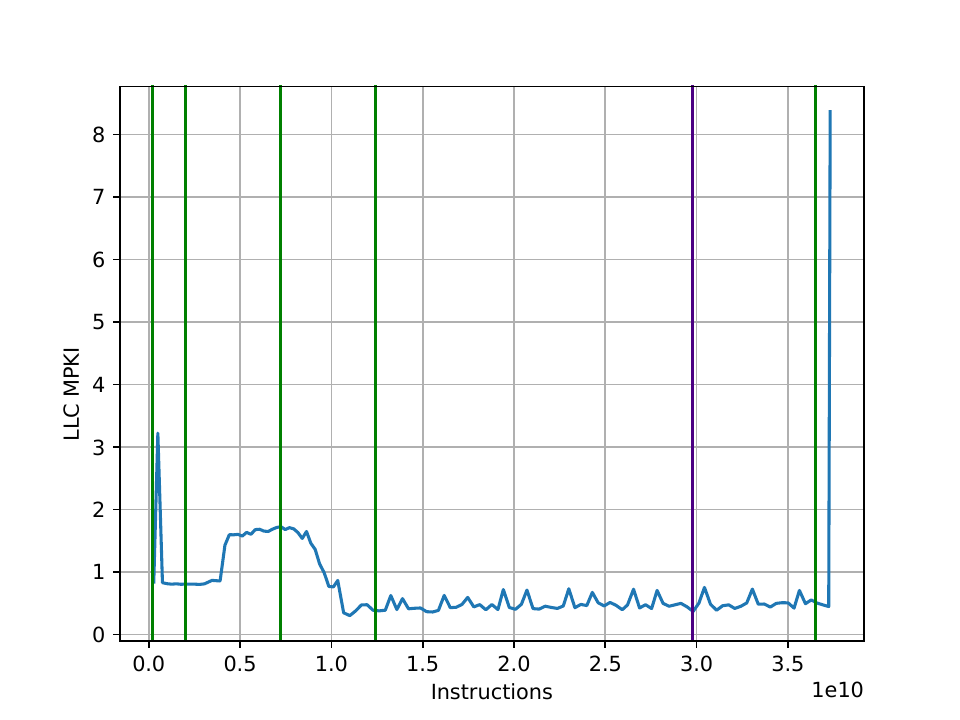}}
\subfigure[gcc with train input]{\includegraphics[width=58mm]{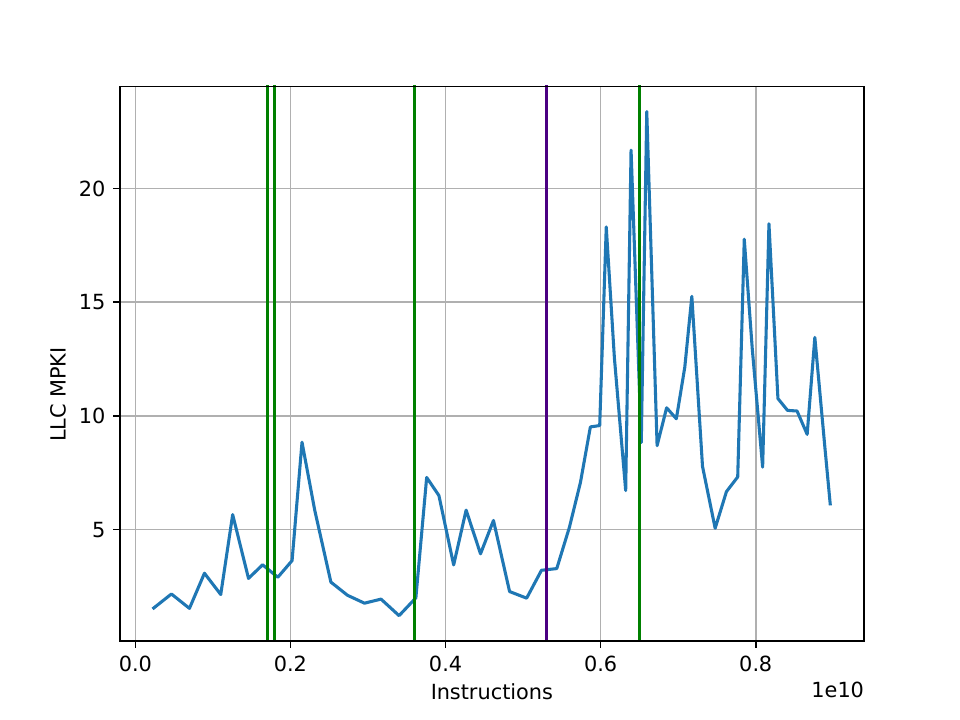}}
\subfigure[mcf]{\includegraphics[width=58mm]{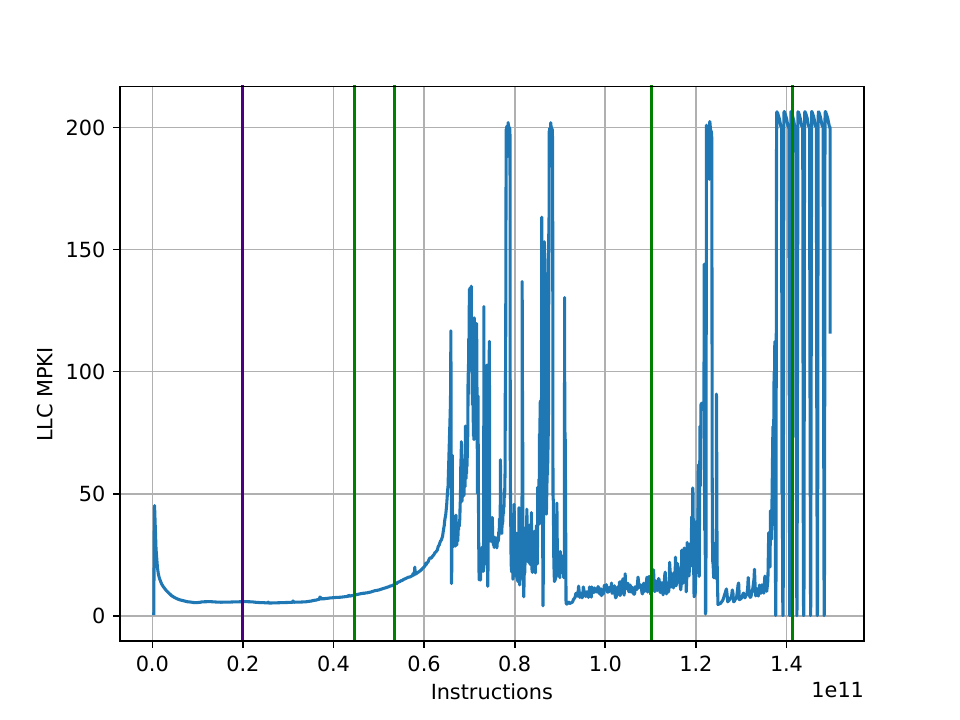}}
\subfigure[omnetpp]{\includegraphics[width=58mm]{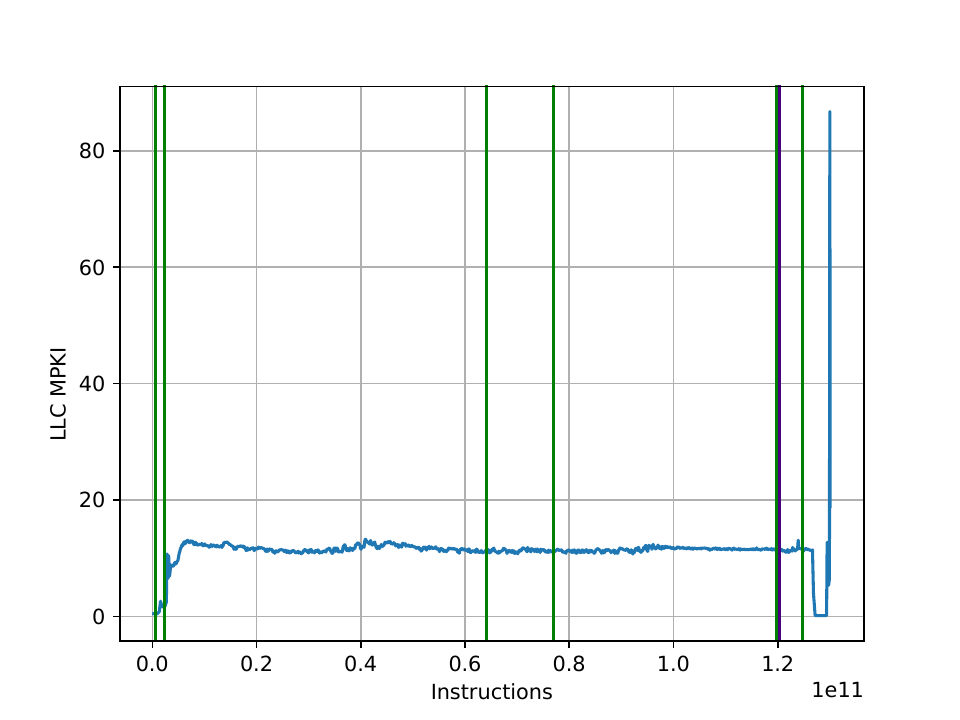}}
\subfigure[xz with IMG 2560 input]{\includegraphics[width=58mm]{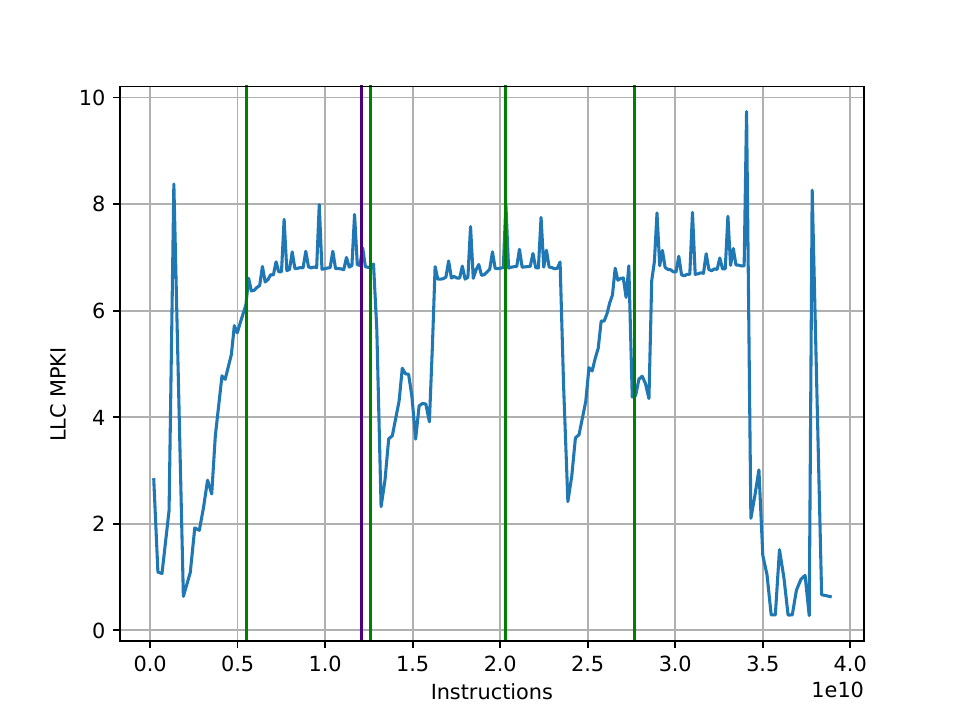}}
\subfigure[wrf]{\includegraphics[width=58mm]{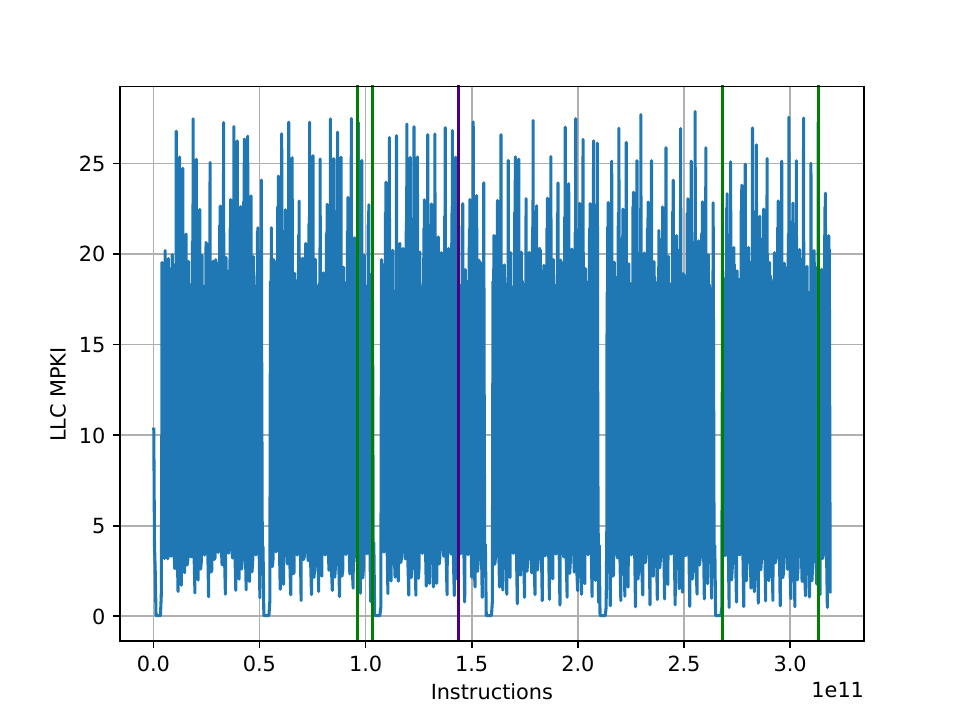}}
\caption{LLC MPKI values obtained with the full simulation, using gem5, for different benchmarks as instructions are executed using LRU policy.} \label{fig:mpkis_whole_bench}
\end{figure*}

As for the CPI values, the figures obtained exhibit the same trends as the MPKI values, with significant differences (and also changes in the relative order between policies) depending on the simulated instruction window. Indeed, the average maximum difference (derived from any of the replacement policies evaluated) between the CPI values using SimPoint and the full simulation is close to 15\%. In applications with high LLC activity, such as \emph{mcf}, in which we are especially interested (MPKI variations can lead to high impact on performance), this variation is up to 40\%.

\begin{framed}
Overall, we can conclude that, in many applications, conventional simulation techniques may affect the insights derived from the evaluation of cache approaches, either because the ordering in the relative performance is changed with respect to that obtained with the full simulation, or because the relative differences among the approaches assessed are significantly far from those obtained when the entire simulation is performed. 
\end{framed}

Moreover, if for a specific metric conventional simulation strategies report values significantly far from those of the full simulation, this can also impact the accuracy on other metrics which depend on the first one, such as energy consumption on the memory hierarchy and the processor, memory endurance or CPI values, which all depend on the LLC MPKI values.  

\subsubsection{LLC values for the whole execution}

Previous results suggest that conventional simulation intervals do not correctly capture the cache behaviour along the entire execution of applications. To further confirm this, we measure how the MPKI value varies during the whole simulation as instructions are executed. This way, we are able to check if the simulation intervals defined by SimPoint are located in code regions with relevant LLC activity (we repeated this experiment with data from PMCs on real execution, obtaining analogous results).

Fig. \ref{fig:mpkis_whole_bench} shows these data for six of the evaluated benchmarks, where the vertical bars show the starting points of the five most-weighted SimPoint intervals for each application (the highest one is highlighted in purple; the other four in green). For the sake of completeness, 
we now show the results for two of the previously evaluated benchmarks (\emph{gcc} and \emph{mcf}), two applications exhibiting changes in the relative order between cache replacement policies in terms of both MPKI and CPI but not previously shown (\emph{perlbench} and \emph{omnetpp}) and two applications that do not experiment these changes in terms of MPKI nor CPI (\emph{xz} and \emph{wrf}).

As illustrated, for \emph{perlbench}, \emph{gcc} \emph{mcf} and \emph{omnetpp} --which all exhibit irregular patterns in LLC activity--, SimPoint intervals do not capture the zones of high LLC activity. Notably, in these benchmarks, the main simulation interval (the only one employed by many authors when they apply the SimPoint technique) does not cover --with a typical simulation window of 100M instructions-- any zone of the code with high LLC activity. Moreover, almost all these SimPoint intervals are located in zones with low LLC activity. 
This is because SimPoint relies on a static criteria based on the code's similarity for selecting the simulation intervals, and this does not necessarily imply a direct correspondence with the LLC behaviour, which is more directly related to the phases of the code. Thus, for \emph{xz} and \emph{wrf}, both of which exhibit a regular LLC activity pattern with highly defined phases, SimPoint intervals do capture the parts of the code with high LLC activity. In fact, with such a regular pattern, regular sampling is likely to report satisfactory results in identifying representative zones of the code in terms of LLC activity.

\subsection{Proposal}
\label{subsec:prop}

\begin{framed}
We just verified that the simulation intervals defined by SimPoint do not capture the zones of high LLC activity, where the replacement policy plays a more important role, so that the various approaches cannot be properly compared under SimPoint or other conventional simulation schemes commonly employed. 
\end{framed}
As a result, we now propose a systematic simulation methodology oriented to employ code fragments that are more representative of the applications' LLC activity and, therefore, to allow a more correct comparison among cache-related proposals. However, it is also needed to maintain a high level of representativeness in terms of performance. To balance both goals and not increase the simulation time required, we suggest employing the same simulation intervals defined by SimPoint, but redefining the associated weights. Essentially, we suggest sorting the intervals based on different criteria related to the LLC activity. Thus, the SimPoint interval exhibiting the highest value according to this criterion becomes the interval with the highest weight in our approach. 

Consequently, aimed to derive from the evaluation of LLC-related proposals, such as cache replacement policies, the same conclusions on the relative performance among them than those of the full simulation, we have experimented with different criteria that assign more representativeness of the overall LLC behaviour to those intervals of the program execution where the LLC suffers a high level of pressure due to significant numbers of MPKI. Accordingly to this goal, we explored different criteria when sorting the simulation intervals for each benchmark, but we focused on the following two approaches as they report the most satisfactory results:

\begin{enumerate}
\item \emph{mpkilru}: The average MPKI obtained in each interval when the LRU policy is employed. The weight of each interval is proportional to its LLC MPKI, so the weight for a specific simulation interval \emph{s} within an application is calculated as follows:

\begin{equation}
weight_{s}= \frac{MPKI_{LRU,s}}{\sum_{i=1}^{n}MPKI_{LRU,i}}
\label{eq1}\end{equation}

where \emph{n} denotes the number of simulations intervals defined by SimPoint for a particular benchmark.

\item \emph{mpkimax}: The maximum LLC MPKI value obtained among all assessed cache replacement policies. Analogously, the weight of each simulation interval within an application is computed as follows: 
\begin{equation}
weight_{s}= \frac{MPKImax_{s}}{\sum_{i=1}^{n}MPKImax_{i}}
\label{eq2}\end{equation}

\end{enumerate}

\vspace{0.7mm}

The different steps of the described simulation methodology are recapped in Algorithm \ref{alg:alg_proposal_mpkilru} for the \emph{mpkilru} approach and in Algorithm \ref{alg:alg_proposal_mpkimax} for the \emph{mpkimax} strategy.

\begin{algorithm}
\caption{Configuration of simulation intervals (mpkilru)}\label{alg:alg_proposal_mpkilru}
\begin{algorithmic}
\STATE {Establish simulation intervals using SimPoint}
\STATE {Weight redefinition according to MPKI LRU}
\REQUIRE{LLC MPKI values for each SimPoint interval with LRU}
\ENSURE{New weights for the intervals defined by SimPoint}
    \FOR{All intervals}
    \STATE Determine the LLC MPKI value with LRU policy and accumulate it in \emph{sum} 
    \ENDFOR
    \FOR{Every single interval s}
        \STATE Divide its LLC MPKI value obtained with LRU policy by \emph{sum}
    \STATE Assign the previous result as new weight
    \ENDFOR

\end{algorithmic}
\end{algorithm}

\begin{algorithm}
\caption{Configuration of simulation intervals (mpkimax)}\label{alg:alg_proposal_mpkimax}
\begin{algorithmic}
\STATE {Establish simulation intervals using SimPoint}
\STATE {Weight redefinition according to MPKI max}
\REQUIRE{LLC MPKI values for each SimPoint interval for all evaluated cache replacement policies}
\ENSURE{New weights for the intervals defined by SimPoint}
\FOR{All intervals}
\STATE Determine the maximum LLC MPKI value among all evaluated cache replacement policies and add it in \emph{sum}
\ENDFOR
\FOR{Every single interval s}
\STATE Divide its maximum LLC MPKI value among all policies by \emph{sum}
\STATE Assign the previous result as new weight
\ENDFOR
\end{algorithmic}
\end{algorithm}

It is worth noting that our \emph{mpkilru} and \emph{mpkimax} approaches do not require any extra simulation time with respect to conventional SimPoint. In the case of original SimPoint, the final value of a particular metric (such as LLC MPKI) is calculated by weighting the metric values obtained in each of the simulation intervals employing the original weights defined by this simulation technique. In our proposals, we employ the same simulation intervals as conventional SimPoint, but changing the associated weights as described in \eqref{eq1} and \eqref{eq2} for \emph{mpkilru} and \emph{mpkimax}, respectively. In the case of \emph{mpkilru} we need to perform the simulation with the LRU policy first, in order to compute then the final LLC MPKI values obtained with the other cache replacement policies by using for the different simulation intervals the weights obtained from the LRU simulation. When \emph{mpkimax} is employed, the only one restriction is that we first need to perform the simulation of all cache replacement policies evaluated, in order to obtain for each interval the maximum LLC MPKI value among all policies (new weights), and therefore to be able to compute the final LLC MPKI values for all the replacement policies. It is also noticeable that when following the original SimPoint simulation strategy using a simulator like gem5, a checkpoint of every region to be simulated is computed first. This allows to replay these regions much faster when performing an architectural exploration. According to our experiments, this time is negligible compared to the checkpoint generation itself (approximately 69 times shorter on average).

Moreover, it is also important to highlight that our approaches, as original SimPoint, entail the simulation of a number of instructions much lower than that of the full simulation. As previously illustrated in Table \ref{tab:number_simpoints}, the number of 100M-instruction simulation intervals we employ ranges between 6 and 28 depending on the benchmark evaluated, so the number of instructions we simulate varies between 600M and 2800M instructions. However, when the applications are entirely executed, the number of instructions simulated is generally significantly higher.  
Table \ref{tab:number_instructions} shows, for all the evaluated benchmark-input pairs, the total number of instructions that the full simulation entails (referred to as \emph{full size} in the table) and the percentage that the instructions simulated with our approaches represent over the \emph{full size} (denoted in the table as \emph{spt vs full}). According to Table \ref{tab:number_instructions}, considering all benchmarks we are simulating, on average, the 2.14\% of the total instructions (again, each benchmark contributes only one value to the final mean --we use the average value for benchmarks with more than one input--), so we can also infer that the simulation time of our approaches is significantly lower than that of the entire simulation of the applications.

\begin{table}
    \caption{Size of the full workloads in number of instructions per evaluated benchmark-input pairs and relative portion (\%) of instructions selected by SimPoint and our approaches}
    \centering
    \begin{tabular}{|c|c|c|c|}
    \hline
     \textbf{Benchmark}   & \textbf{Input} & \textbf{Full size} & \textbf{spt vs full}\\
     \hline
     \multirow{5}{4em}{perlbench }  & diffmail & 7,66x$10^{10}$ & 1,57\\
     \cline{2-4}
         & perfect & 1,24x$10^{10}$ & 4,83\\
         \cline{2-4}
         & scrabbl & 2,94x$10^{10}$ & 3,06\\
         \cline{2-4}
         & splitmail & 3,73x$10^{10}$ & 3,48\\
         \cline{2-4}
         & suns & 1,50x$10^{9}$ & 99,71 \\
    \hline
    \multirow{3}{2em}{gcc}   & 200 & 1,06x$10^{11}$ & 1,80 \\
    \cline{2-4}
         & scilab & 8,10x$10^{10}$ & 2,59\\
         \cline{2-4}
         & train & 8,96x$10^{9}$ & 7,81\\
    \hline
    \multirow{2}{3.5em}{bwaves}   & bwaves1 & 7,36x$10^{11}$ & 0,20 \\
    \cline{2-4}
         & bwaves2 & 7,68x$10^{11}$ & 0,18\\
    \hline
    mcf    & train & 1,50x$10^{11}$ & 1,54\\
    \hline
    cactuBSSN    & train & 1,65x$10^{11}$ & 1,21\\
   \hline
   lbm    & train & 8,16x$10^{11}$ & 0,33\\
   \hline
   omnetpp    & train & 1,30x$10^{11}$ & 1,92\\
   \hline
   wrf    & train & 3,19x$10^{11}$  & 0,78\\
   \hline
   xalancbmk    & train & 2,93x$10^{11}$ & 0,92\\
   \hline
   x264    & train & 2,51x$10^{11}$ & 0,52\\
   \hline
   cam4    & train & 7,47x$10^{11}$ & 0,38\\
   \hline
   pop2    & train & 5,68x$10^{11}$ & 0,35\\
   \hline
   deepsjeng    & train & 3,51x$10^{11}$ & 0,34\\
   \hline
   imagick    & train & 2,94x$10^{11}$ & 0,75\\
   \hline
   leela    & train & 3,70x$10^{11}$ & 0,43\\
    \hline
    \multirow{2}{1.75em}{nab }   & aminos & 6,80x$10^{10}$ & 3,38 \\
    \cline{2-4}
         & gcn4dna & 6,54x$10^{11}$ & 0,26\\
    \hline
    exchange2    & train & 3,64x$10^{11}$ & 0,44\\
    \hline
    fotonik3d    & train & 2,00x$10^{11}$ & 1,05 \\
    \hline
    roms    & train & 1,07x$10^{12}$ & 0,21\\
    \hline
    \multirow{2}{1.5em}{xz }   & input\_combined 40 & 8,86x$10^{10}$ & 1,81 \\
    \cline{2-4}
         & IMG\_2560 40 & 3,89x$10^{10}$ & 4,12\\
    \hline
    \end{tabular}
    \label{tab:number_instructions}
\end{table}

Finally, please also note that it is expected that analogous criteria to the \emph{mpkilru} and \emph{mpkimax} approaches proposed for cache replacement policies could be applied when other type of cache-related proposals are evaluated.

%% file: evaluation.tex
\section{Evaluation}
\label{sec:evaluation}

In this section we assess our proposals by employing two metrics: \emph{order} and \emph{closeness}, which are discussed next.
 
\subsection{Relative ordering of replacement policies from LLC MPKI results}

Our goal is to obtain the same relative order among the replacement policies used (in terms of LLC MPKI) from the simulation intervals as that observed in the full simulation. Thus, we measure how close of this goal we are by comparing the relative order experienced --in our proposals vs. the entire simulation-- by each pair of employed cache replacement policies. Although we evaluate five approaches in our experiments, we do not take into account the \emph{random} results due to their unpredictable behaviour, so we work with six pairs (combinations 1-2, 1-3, 1-4, 2-3, 2-4 and 3-4; each number identifies one of the four cache policies used).

Our proposed per-benchmark metric named \emph{order} (initially set to zero) ranges between 0 and 6 and is computed as follows: if a specific pair maintains the same relative order under our simulation intervals as when simulating the entire benchmark, the metric remains unchanged; otherwise, the metric is incremented by one. Thus, an application that under our approach exactly matches the same relative order between the four cache replacement policies derives an \emph{order} value of 0, and if it provides different orders between the six evaluated pairs it reports an \emph{order} of 6. 
Thus, numbers close to zero indicate a high level of similarity with the full simulation. 

Table \ref{tab:order} recaps the average \emph{order} obtained using just the original most-weighted SimPoint (denoted as \emph{weight}), using \textbf{all} SimPoint intervals with original weights (\emph{spt} approach), and also using our \emph{mpkilru} and \emph{mpkimax} proposals, where we employ \textbf{all} the SimPoint intervals but ordered and weighted as stated in Section~\ref{subsec:prop}.  
We show \emph{order} values in four different scenarios:
\begin{itemize}
\item Considering all the 20 evaluated benchmarks (\emph{Avg}). 
\item Excluding the applications with MPKI values --under LRU policy-- below 0.1 (\emph{Avg w/o low}). 
\item Considering only the seven applications with the highest MPKIs (\emph{Avg-high}). Note that we have chosen the number of benchmarks needed to obtain an accumulated MPKI value --under the LRU replacement scheme-- which exceeds the 80\% of the total accumulated MPKI number considering all the 20 evaluated benchmarks. 
\item Considering only those benchmarks where the relative order among cache policies in the \emph{spt} approach does not match the order of the full simulation (\emph{Avg-changes}). Note that for all programs where the relative order between policies obtained with \emph{spt} matches that of the full simulation, our two proposals manage to report the same order as well. 
\end{itemize}

\begin{table}
\caption{Order metric for different proposals\label{tab:order}}
\begin{center}{\tt
    \begin{tabular}{|l|c|c|c|c|}
    \hline
    {}&\textbf{weight}&\textbf{spt}&\textbf{mpkilru}&\textbf{mpkimax}\\
    \hline
    Avg & 2.57 & 1.14 & 0.79 & 0.86\\
    \hline
    Avg w/o low & 2.47 & 1.11 & 0.73 & 0.81\\
    \hline
    Avg-high & 3.19 & 1.79 & 1.31 & 1.26\\
    \hline
    Avg-changes & 3.45 & 1.89 & 1.31 & 1.39\\
    \hline
    \end{tabular}}
\end{center}
\end{table}

As illustrated, our proposals report the lowest \emph{order} value in all the four scenarios assessed. 

\begin{framed}
Overall, \emph{mpkilru} reduces the number of changes in the relative order among the policies by more than 30\% with respect to conventional SimPoint (\emph{spt}). 
\end{framed}

The same reduction is also achieved when considering only the benchmarks exhibiting changes in \emph{spt} with respect to the full simulation order. 
In the case of memory-intensive applications, our \emph{mpkimax} and \emph{mpkilru} schemes report \emph{order} values 30\% and 27\% lower than that of \emph{spt}, respectively. Furthermore, we do demonstrate that using only one SimPoint with its original weight (an approach used by many authors), significantly increases the number of changes in the relative order among the policies, leading to incorrect comparisons between cache replacement schemes as previously explained in Section \ref{subsec:mot}. 

If we focus on individual applications, we may highlight benchmarks such as \emph{x264} and \emph{pop2}, which with \emph{spt} exhibit \emph{order} values of 4 and 3, respectively (therefore a relative order among the replacement policies assessed quite far from that of the full simulation), but that when using our proposals they derive \emph{order} values of 1 and zero, respectively, so that they practically match the same behaviour as when the entire simulation is performed.

\begin{figure*}[htbp]
\centering
\subfigure[gcc with 200 input]{\includegraphics[width=58mm]{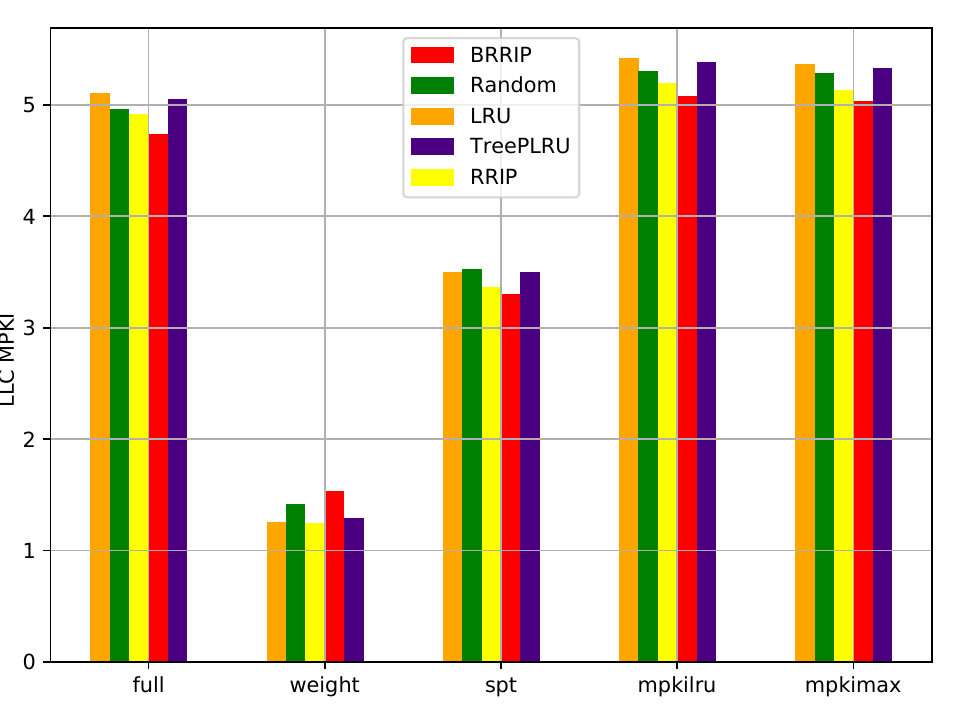}}
\subfigure[mcf]{\includegraphics[width=58mm]{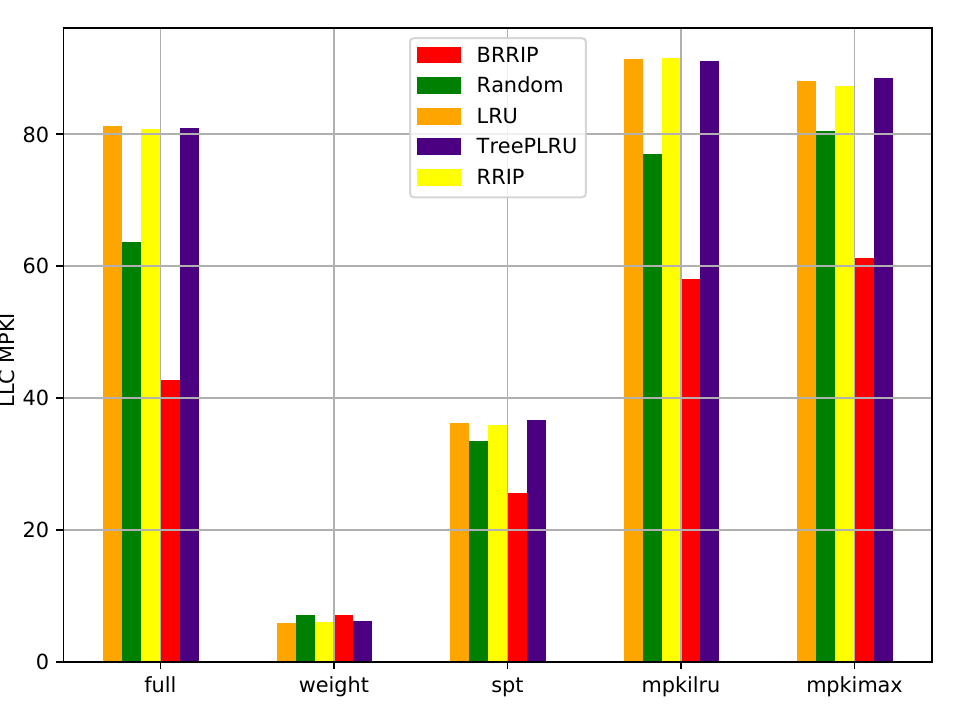}}
\subfigure[roms]{\includegraphics[width=58mm]{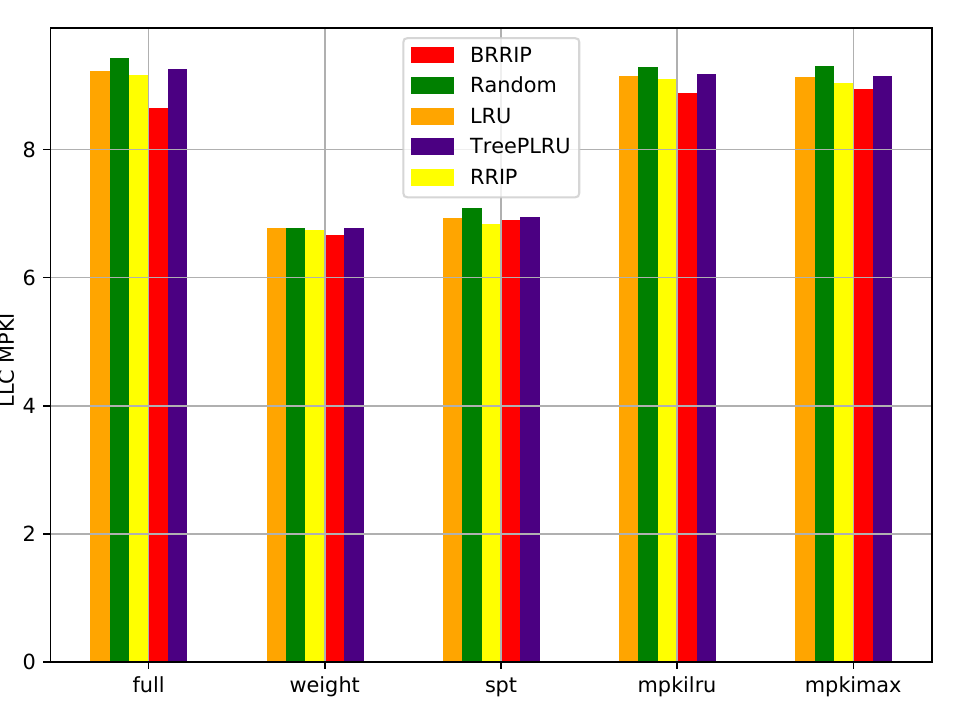}}
\subfigure[wrf]{\includegraphics[width=58mm]{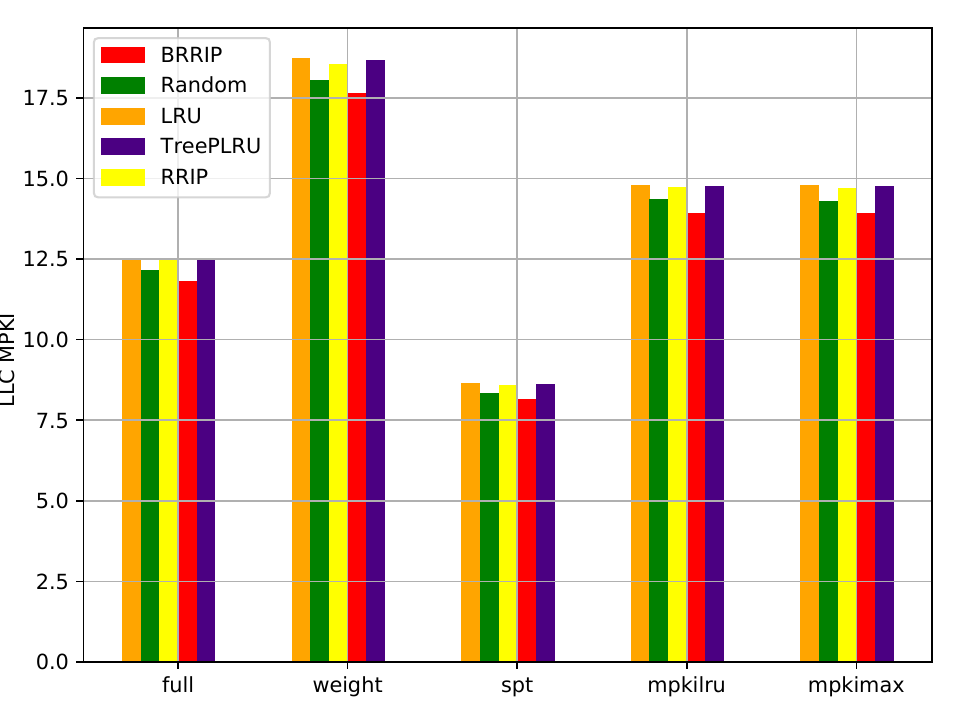}}
\subfigure[x264]{\includegraphics[width=58mm]{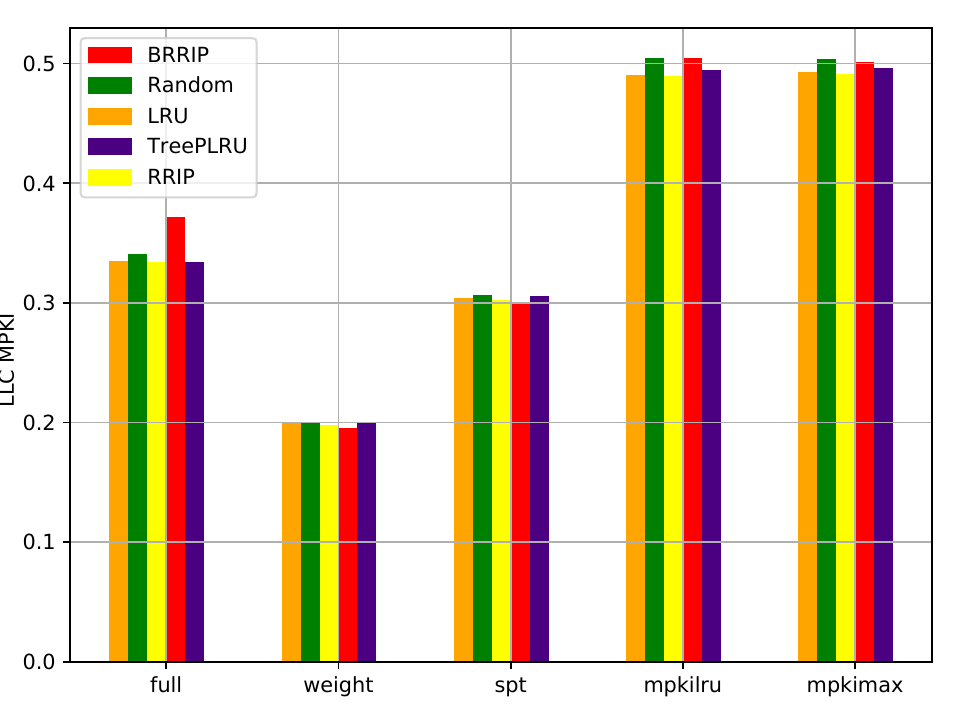}}
\subfigure[leela]{\includegraphics[width=58mm]{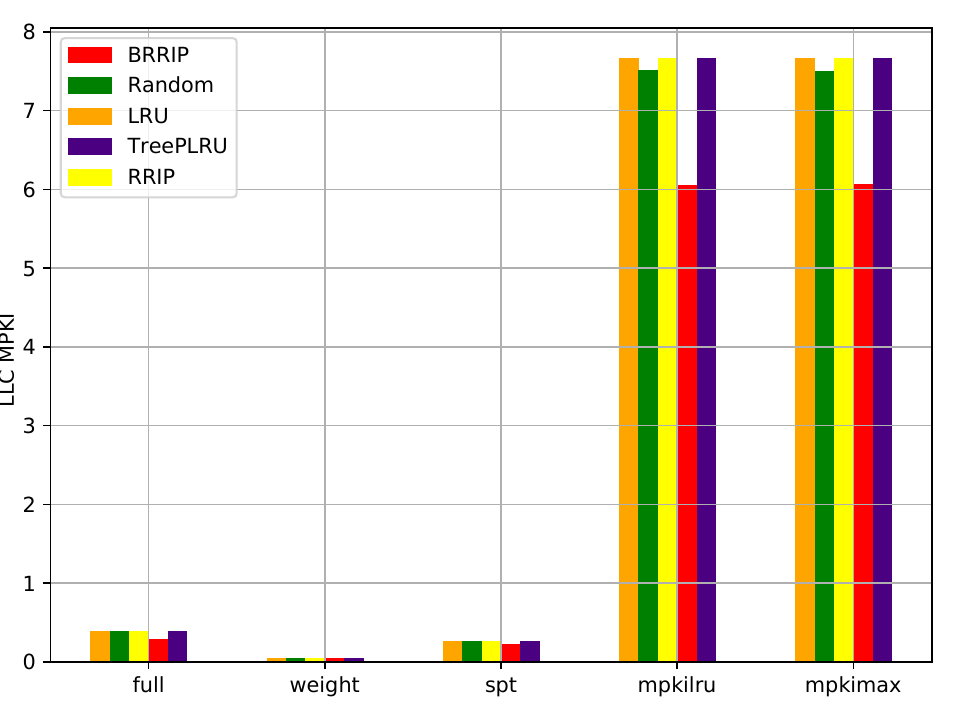}}
\caption{LLC MPKI obtained with full simulation, \emph{weight}, \emph{spt}, \emph{mpkilru} and \emph{mpkimax} simulation strategies, using gem5, for different benchmarks and cache replacement policies.} \label{fig:mpkis_proposals}
\end{figure*}

\subsection{Closeness to absolute full simulation MPKI numbers }
Although our proposals have been demonstrated to provide a higher level of similarity with the full simulation order than that of conventional SimPoint, we also pursue the goal of reporting MPKI values close to those of the full simulation because, as previously stated, these numbers are usually underestimated in conventional simulation schemes. 

Fig. \ref{fig:mpkis_proposals} illustrates --for six benchmarks as a representative sample of the results obtained-- the LLC MPKI values derived under the \emph{spt}, \emph{weight}, \emph{mpkilru} and \emph{mpkimax} approaches  
as well as by the full simulation. We report the results for four applications also shown in the motivational study (\emph{gcc}, \emph{mcf}, \emph{roms} and \emph{x264}) and two other benchmarks (\emph{wrf} and \emph{leela}). For the three applications in the upper part of Fig. \ref{fig:mpkis_proposals}, we observe that both of our proposals report MPKI numbers much closer to those of the full simulation than the original SimPoint (both \emph{weight} and \emph{spt} alternatives). In the case of \emph{wrf}, our proposals also obtain LLC MPKI values closer to those of the entire simulation than conventional SimPoint, overestimating the values of the full simulation moderately less than the \emph{spt} underestimate them. In the \emph{x264} benchmark, \emph{spt} can report MPKI values closer to those of the full simulation than our proposals, but it does not capture the high MPKI value of BRRIP (compared to the other policies), which do capture both of our proposals. Finally, for the \emph{leela} application (and also in the case of \emph{cam4}, not shown in the graph), our proposals significantly overestimate the MPKI values, leading to numbers notably far from those of the full simulation and \emph{spt}. To quantify the closeness of the MPKI values reported by the various simulation approaches from the values derived from the full simulation, we introduce the lower-is-better \emph{closeness} metric. For each specific simulation strategy, this metric accumulates the percentage deviation of the LLC MPKI values obtained for all evaluated replacement policies (except \emph{random}) from those obtained with the entire simulation. Hence, low values of \emph{closeness} under a certain simulation approach imply that it is more accurate to reproduce MPKIs derived from the full simulation, and therefore it is also more likely to obtain the same conclusions from the evaluation of LLC-related proposals employing the specific simulation intervals as when performing the entire simulation. Accordingly, we define the metric as follows:

\begin{equation}
closeness(MPKI)=\sum_{i=1}^{4}|\frac{MPKI_{i,full}-MPKI_{i,proposal}}{MPKI_{i,full}}|
\label{eq3}\end{equation}

\begin{table*}[tbp]
\caption{MPKI closeness metric (arithmetic and geometric mean) for different simulation proposals\label{tab:closeness_mpki}}
\begin{center}{\tt
    \begin{tabular}{l|c|c|c|c||c|c|c|c|}
\cmidrule{2-9} \\ [-0.42cm]
 & \multicolumn{4}{c||}{\textbf{Arithmetic mean}} & \multicolumn{4}{c|}{\textbf{Geometric mean}} \\
    \hline
{}&\textbf{weight}&\textbf{spt}&\textbf{mpkilru}&\textbf{mpkimax}&\textbf{weight}&\textbf{spt}&\textbf{mpkilru}&\textbf{mpkimax}\\
    \hline
    Avg w/o low & 1.85 & 0.91 & 6.65 & 6.62 & 1.18 & 0.58 & 1.28 & 1.28\\
    \hline
    Avg w/o low +2 & 1.67 & 0.94 & 1.54 & 1.51 & 1.04 & 0.59 & 0.84 & 0.84\\
    \hline
    Avg-high & 1.36 & 0.87 & 0.68 & 0.66 & 0.59 & 0.51 & 0.32 & 0.33\\
    \hline
    \end{tabular}}
\end{center}
\end{table*}

Table \ref{tab:closeness_mpki} shows the arithmetic and geometric means of MPKI \emph{closeness} obtained 
in the \emph{Avg w/o low} and \emph{Avg-high} scenarios already considered for the \emph{order} metric, as well as when considering all applications except 
those with MPKI values below 0.1 and the outliers \emph{leela} and \emph{cam4} (\emph{Avg w/o low +2)}. We do not show results when considering all applications because the \emph{closeness} metric in applications with very low MPKI, such as \emph{exchange2} (on the order of $10^{-5}$), reaches extraordinarily high and distorting values (higher than 5000 for all simulation strategies). This is why we also shown geometric mean values in order to mitigate the effect of the extraordinary contribution to the final arithmetic mean value of a few applications with low values of LLC MPKI. In addition, we do not report the results for the (\emph{Avg-changes}) scenario because it only makes sense in the context of the \emph{order} metric. 

As shown, our proposals report \emph{closeness} values significantly higher than those of original SimPoint when considering all applications except those with MPKI values below 0.1 (\emph{Avg w/o low} scenario) and we employ the arithmetic mean of the values reported by the different benchmarks (this difference significantly decreases when the geometric mean is used).  This is related to applications exhibiting moderately low MPKI values (much less relevant when comparing cache-related approaches such as cache replacement policies), since small variations in absolute MPKI numbers may still lead to high \emph{closeness} numbers. This is the case of \emph{x264} and \emph{leela} applications shown in Fig. \ref{fig:mpkis_proposals}, which according to Table \ref{tab:number_simpoints}, exhibit LLC MPKI numbers with the LRU policy of just 0.34 and 0.38, respectively. Note also that for the \emph{cam4} application, not being a benchmark exhibiting low LLC MPKI values, our approaches significantly overestimate this metric. In this case it has to do with the fact that the new weights that our strategies assign to the 28 SimPoint intervals of this program present a significant imbalance (due to the high disparity in LLC MPKI values across SimPoint intervals), much greater than in most of the remaining applications. This application has the highest number of simulation intervals of all evaluated programs, increasing the probability that some intervals capture zones of high LLC activity. Although original most-weighted simulation intervals in \emph{cam4} do not fall in this kind of zones, there are other original low-weighted intervals that do capture these zones. Notably, the simulation interval exhibiting the second lowest weight according to original weights (lower than 0.1\%) becomes the most-weighted interval in our approaches, contributing to the final LLC MPKI value with more than 30\%, significantly more than any of the other intervals, reporting an LLC MPKI value just for this interval around 4X the mean value obtained with the full simulation. If we focus on the four most-weighted intervals in our \emph{mpkilru} approach, they contribute with more than 56\% to the final LLC MPKI value, while the same four intervals using their original weights contribute to the final value with just 2.2\%. It is also important to recall that as our \emph{mpkilru} and \emph{mpkimax} approaches are assigning higher weights to those simulation intervals with high LLC activity (high LLC MPKI numbers), it was expected that our strategies overestimate MPKI numbers in some cases. However, the differences with respect to the full simulation numbers clearly decrease when we also exclude the outliers \emph{cam4} and \emph{leela} applications. 
Moreover, when we only consider the seven most memory-intensive programs, our \emph{mpkilru} and \emph{mpkimax} proposals manage to outperform all the other simulation strategies, reporting MPKI \emph{closeness} values when we employ the arithmetic mean approximately 22\% and 24\% lower than that of \emph{spt}, respectively, and around 37\% and 35\% respectively when the geometric mean is used. We consider this as significantly relevant, as our approaches are especially targeted to evaluate microarchitectural proposals involving the cache system, such as LLC replacement policies, which play a more important role in applications with high LLC MPKI numbers.

\begin{framed}
Hence, for memory-intensive applications and compared to SimPoint, we significantly improve the representativeness of our simulation intervals --in terms of LLC MPKI numbers-- with respect to the full simulation, 
\end{framed}

as illustrated in Fig. \ref{fig:mpkis_proposals} for \emph{mcf}, \emph{roms}, \emph{gcc} and also \emph{wrf}, all of them applications with high LLC MPKI numbers.

\begin{table*}[tbp]
\caption{CPI closeness metric (arithmetic and geometric mean) for different simulation proposals\label{tab:closeness_cpi}}
\begin{center}{\tt
    \begin{tabular}{l|c|c|c|c||c|c|c|c|}
\cmidrule{2-9} \\ [-0.42cm]
 & \multicolumn{4}{c||}{\textbf{Arithmetic mean}} & \multicolumn{4}{c|}{\textbf{Geometric mean}} \\
    \hline
{}&\textbf{weight}&\textbf{spt}&\textbf{mpkilru}&\textbf{mpkimax}&\textbf{weight}&\textbf{spt}&\textbf{mpkilru}&\textbf{mpkimax}\\
    \hline
    Avg & 0.86 & 0.56 & 1.44 & 1.43 & 0.34 & 0.16 & 0.37 & 0.39\\
    \hline
    Avg w/o low & 0.95 & 0.62 & 1.39 & 1.37 & 0.47 & 0.20 & 0.36 & 0.37\\
    \hline
    Avg w/o low +2 & 1.00 & 0.69 & 0.86 & 0.86 & 0.51 & 0.25 & 0.26 & 0.27\\
    \hline
    Avg-high & 0.79 & 0.53 & 0.46 & 0.45 & 0.36 & 0.34 & 0.13 & 0.15\\
    \hline
    \end{tabular}}
\end{center}
\end{table*}

\subsection{Closeness to absolute full simulation CPI numbers }
Although we have demonstrated that our proposed simulation strategy outperforms the conventional SimPoint in terms of the \emph{order} metric in all scenarios evaluated, as well as in terms of MPKI \emph{closeness} in the case of memory-intensive programs, we must also explore how our proposals work in terms of performance if these approaches aim to postulate as an alternative to conventional simulation schemes. For this purpose, we also introduce the CPI \emph{closeness} metric, which is defined --analogously to the case of MPKI-- as follows:

\begin{equation}
closeness(CPI)=\sum_{i=1}^{4}|\frac{CPI_{i,full}-CPI_{i,proposal}}{CPI_{i,full}}|
\label{eq4}\end{equation}

Table \ref{tab:closeness_cpi} recaps the geometric and arithmetic means of the \emph{closeness} obtained for CPI values in the same scenarios as in the case of the LLC MPKI \emph{closeness}, and also when considering all the benchmarks evaluated. As expected, \emph{spt} reports the CPI values closest to those of the full simulation when considering all applications. In the same scenario, when we employ the geometric mean, our \emph{mpkilru} and \emph{mpkimax} approaches are able to report CPI values significantly close to those of the \emph{weight} approach, although still moderately far from \emph{spt}. This could be considered as expectable, as we are mainly focusing on LLC activity to assign weights to SimPoint intervals and it is important to note that original SimPoint defines simulation intervals and the corresponding weights aimed to reproduce the overall behaviour of applications mainly in terms of performance. However, the differences between \emph{spt} and our proposals decrease when we do not take into account the programs with low LLC activity, until the point where our proposals practically match (especially when considering the geometric mean) the performance numbers reported by \emph{spt} when we exclude the applications with LLC MPKI values below 0.1 and also the outliers \emph{leela} and \emph{cam4} (\emph{Avg w/o low +2)} scenario. More importantly, even when we just consider the most memory-intensive applications, our proposals manage to report performance \emph{closeness} values around 13-15\% lower than that of SimPoint when the arithmetic mean is used and around a significant 56-62\% when we employ the geometric mean, with CPI numbers significantly close to those of the full simulation. In this way, with our redefinition of the weights associated to the simulation intervals defined by original SimPoint, we effectively achieve a satisfactory trade-off in reproducing the overall behaviour of applications in terms of LLC activity and also performance. As a result, for programs with high LLC MPKI numbers we significantly outperform \emph{spt} in terms of MPKI \emph{closeness} (as expected and previously shown) but also in terms of CPI \emph{closeness}, despite of being original SimPoint a technique mainly conceived to match performance numbers of full execution. This also reveals that the impact of an accurate determination of LLC MPKI on other metrics such as CPI is significantly relevant for memory-intensive programs, where the LLC activity is high, so that original SimPoint is, generally, increasingly less accurate on CPI values as we progressively consider only more memory-intensive applications (see \emph{spt} column from top to bottom in Table \ref{tab:closeness_cpi} in the case of the geometric mean) whereas our approaches follow exactly the opposite trend.

\begin{framed}
We can conclude that for memory-intensive benchmarks, our simulation intervals obtain, also in terms of performance numbers, a higher level of representativeness of the entire simulation than original SimPoint. 
\end{framed}

%% file: conclusions.tex
\section{Conclusions}
\label{sec:conclusions}

In this paper, we first demonstrated our hypothesis regarding the evaluation of microarchitectural cache-related proposals: the particular simulation window employed can lead to incorrect conclusions. As a motivational case study, we explored the impact of different commonly used simulation windows on the performance of various replacement policies implemented in the LLC. This analysis made it possible to infer that current simulation strategies do not fully capture the behaviour of the LLC; therefore the specific simulation window employed may entail wrongful comparisons.

Consequently, we also proposed a different simulation strategy oriented to maintain a proper trade-off in reproducing the overall behaviour of applications in terms of both LLC activity and performance, without affecting the simulation time. 
For this purpose, we suggested employing the same simulation intervals as SimPoint, but ordered and weighted according to different metrics that take into account the number of LLC misses, aimed to improve the representativeness of the simulation windows for the cache system. 

Our experimental evaluation demonstrated that our approaches outperform conventional SimPoint in terms of the \emph{order} metric (up to 30\%) in all scenarios evaluated, and, in the case of memory-intensive programs, also in terms of MPKI and CPI \emph{closeness} (up to 24 and 15\%, respectively). Overall, we can conclude that our simulation strategies report a satisfactory trade-off in reproducing the overall behaviour of the applications in terms of both LLC activity and performance, particularly in the case of memory-intensive benchmarks, which also makes it possible a more accurate simulation in terms of other features at the whole processor level which depend on the mentioned metrics, such as energy consumption or memory endurance.